\documentclass[letterpaper,12pt]{article}

\usepackage{amssymb,latexsym,amsmath}

\usepackage{fullpage}
\usepackage{bbm} 

\usepackage[pdftex]{graphicx}

\makeatletter \@addtoreset{equation}{section}
\makeatother

\newcommand{\be}{\begin{equation}}
\newcommand{\ee}{\end{equation}}
\newcommand{\bea}{\begin{eqnarray}}
\newcommand{\eea}{\end{eqnarray}}
\newcommand{\nn}{\nonumber}
\newcommand{\ul}{\underline}
\newcommand\bbone{\ensuremath{\mathbbm{1}}}

\newcommand{\vct}{X}
\newcommand{\vielb}{v}

\begin{document}

\begin{titlepage}
	\thispagestyle{empty}
	\begin{flushright}
		\hfill{DFPD-12/TH/11\\KCL-MTH-12-11}
	\end{flushright}
	
	\vspace{35pt}
	
	\begin{center}
	    {\Large \bf{BPS domain walls in $\mathcal N=4$ supergravity and dual flows}}
		\vspace{25pt}
		
		{Davide~Cassani$^{a}$, Gianguido~Dall'Agata$^{b,c}$ and Anton F.~Faedo$^{d}$}
		\vspace{25pt}
		
		{\small
		
		{\it ${}^a$ Department of Mathematics, King's College London,\\
	     The Strand, London WC2R 2LS, United Kingdom}

	    {\tt davide.cassani@kcl.ac.uk}

	    \vspace{12pt}

		{\it ${}^b$ Dipartimento di Fisica e Astronomia ``Galileo Galilei''\\
		Universit\`a di Padova, Via Marzolo 8, 35131 Padova, Italy}\\
		
		\vspace{12pt}
		
		{\it \it ${}^c$  INFN, Sezione di Padova \\
		Via Marzolo 8, 35131 Padova, Italy}\\
		{\tt dallagat@pd.infn.it}		
		
	    \vspace{12pt}

		{\it ${}^d$ Department of Physics, Swansea University,\\
Singleton Park, Swansea SA2 8PP, United Kingdom}\\
{\tt a.f.faedo@swansea.ac.uk}
		}
		\vspace{40pt}
		
		{\bf Abstract}
	\end{center}

\begin{quote}
We establish the conditions for supersymmetric domain wall solutions to $\mathcal N=4$ gauged supergravity in five dimensions. These read as BPS first-order equations for the  warp factor and the scalar fields, driven by a superpotential and supplemented by a set of constraints that we specify in detail.
Then we apply our results to certain consistent truncations of IIB supergravity, thus exploring their dual field theory renormalization group flows. We find a universal flow deforming superconformal theories on D3-branes at Calabi--Yau cones. Moreover, we obtain a superpotential for the solution corresponding to the baryonic branch of the Klebanov--Strassler theory, as well as the superpotential for the flow describing D3 and wrapped D5-branes on the resolved conifold.
\end{quote}		
	  	
	\vspace{10pt}

\end{titlepage}

\baselineskip 6 mm

\tableofcontents

\section{Introduction}

According to the gauge/gravity correspondence, domain wall solutions in (super)gravity theories describe dual renormalization group (RG) flows in which the extra radial coordinate plays the role of the energy scale. Some important early examples for the development of these concepts include \cite{Girardello, FGPW}. In these references, the domain wall interpolates between two AdS solutions and is dual to an RG flow between two different conformal field theories. More sophisticated flows, displaying remarkable properties like confinement and chiral symmetry breaking, were constructed based on the conifold geometry \cite{KlebanovTseytlin, KlebanovStrassler, MaldacenaNunez, PandoZayasTseytlin, BaryonicBranchKS}.

In this paper we will develop a systematic approach to this important class of solutions in five-dimensional, $\mathcal{N}=4$ (i.e.\ half-maximal) gauged supergravity. For our purposes, a (flat) domain wall will be a gravitational solution that preserves four-dimensional Poincar\'e invariance. The general form of the metric thus reads
\begin{equation}\label{dwmetric}
	ds^2 \,=\, e^{2A(r)} ds^2(\mathbb{R}^{1,3}) + dr^2
\end{equation} 
where $A$ is a warp factor, function of the radial coordinate~$r$. These metrics are typically supported by a set of scalar fields $\phi^x$ with a radial profile. An important subclass of domain walls is obtained when the scalar potential can be written, in a very specific way, in terms of a real function, $W(\phi)$, known as the superpotential. In this case, the second-order equations of motion can be solved by imposing a BPS-like condition, which ensures gravitational stability and reads as first-order equations for the radial evolution of the domain wall \cite{S&T}:
\be
A' =   W\,,\qquad\qquad \phi^{x\,\prime} = -3 \, g^{xy}\partial_y W\,,
\ee
where $g_{xy}$ is the scalar kinetic matrix. The superpotential itself is a noteworthy quantity: it defines the tension of the domain wall and provides a monotonic $c$-function for the dual RG flow.
The conditions above emerge naturally when supersymmetry is present: the superpotential and the first-order flow equations arise from the vanishing of the fermionic susy variations. Supergravity provides thus a natural setup to look for them \cite{DWReview}. Nevertheless, given the scalar potential, a superpotential $W$ reproducing it is not necessarily unique nor related a priori with supersymmetry.

An analysis of supersymmetric domain walls in five-dimensional $\mathcal N=2$ supergravity was done some time ago in \cite{CveticDW,5dN=2domainwalls}. In \cite{5dN=2domainwalls} it was found that in addition to the first-order equations driven by a superpotential, supersymmetry imposes an algebraic constraint on the scalar fields. These results were recently applied in \cite{HalmagyiLiuSzepietowski} to study domain walls in the $\mathcal N=2$ truncations of \cite{T11reduction}, lifting to type IIB solutions based on the conifold geometry.
Nonetheless, the various conifold solutions are contained in different $\mathcal N=2$ truncations, and the one describing the baryonic branch of the Klebanov--Strassler theory, found in~\cite{BaryonicBranchKS} building on a proposal of~\cite{GubserHerzogKlebanov}, falls outside the setup of \cite{HalmagyiLiuSzepietowski}. However, a larger consistent truncation has been established that preserves $\mathcal N=4$ supersymmetry and includes all known conifold solutions \cite{T11reduction, Bena:2010pr}. This $\mathcal N=4$ theory thus provides the proper framework for a unified, systematic study of the supersymmetry properties of those solutions. Besides giving a genuine supergravity origin to the known superpotentials, working in this setup we will provide a superpotential for the solution of~\cite{BaryonicBranchKS}, clarifying an issue recently put forward in the literature~\cite{HalmagyiLiuSzepietowski, Giecold}.

The conifold solutions preserve four supercharges, and as we will see 1/4 BPS domain walls emerge naturally in $\mathcal N=4$ gauged supergravity. Our general analysis is inspired by~\cite{5dN=2domainwalls}, but is considerably more involved, since in our case the R-symmetry is USp(4) rather than just SU(2), and three-quarters of the supersymmetry is broken.

The paper is divided in two distinct parts. In the first we analyze the construction of supersymmetric domain walls in five-dimensional $\mathcal N=4$ supergravity with arbitrary matter coupling and very general gauging (the only assumption on the gauging being that the scalar in the gravity multiplet remains neutral). As we summarize in section~\ref{NecSufCond}, we find that the necessary and sufficient conditions for supersymmetry to be preserved take the form of first-order flow equations for the warp factor and the scalar fields, generated by a superpotential computed as an eigenvalue of the gravitino shift matrix. In addition, we obtain a set of algebraic constraints involving the scalars. These are crucial to satisfy the supersymmetry conditions, and take a more complicated form than the one found in \cite{5dN=2domainwalls}. Moreover, the $\mathcal N=4$ supersymmetry parameter is constrained in such a way that four supercharges are generically preserved. We also briefly address the special case in which supersymmetry is enhanced so that eight supercharges are preserved (see~\cite{Zagermann} for previous results on such 1/2 BPS domain walls). 

In the second part, we focus on two distinguished $\mathcal N=4$ supergravity models. These are significant because they can be obtained as consistent truncations of type IIB supergravity, meaning that any solution to the five-dimensional theory is automatically a solution to the ten-dimensional one. The first model arises from dimensional reduction on any manifold admitting a Sasaki--Einstein structure as described in \cite{IIBonSE, GauntlettVarelaIIBonSE}. By implementing our BPS conditions, we provide its most general supersymmetric domain wall solution. This describes universal deformations of the superconformal field theories on D3-branes at the tip of Calabi--Yau cones. Tuning a parameter corresponding in type IIB to an imaginary self-dual three-form flux, this new flow interpolates between the solution in \cite{PandoZayas:2001iw, BenvenutiEtAl} and the one in \cite{GPPZflow}.

The second model is an extension of the preceding one when the base of the conifold is chosen as internal Sasaki--Einstein geometry. Its coset structure permits an SU(2)$\times$SU(2) left-invariant reduction that enlarges the consistent truncation to include a new vector multiplet along with new gaugings \cite{T11reduction, Bena:2010pr}. One can see this truncation as an $\mathcal{N}=4$ supersymmetrization of the Papadopoulos--Tseytlin (PT) ansatz \cite{PapadopoulosTseytlin}, which encompasses all the celebrated conifold solutions. 
Again, by implementing the general conditions found in the first part of the paper, and exploiting the $\mathcal N=4$ structure identified in \cite{T11reduction}, we study the domain walls within the PT subsector of the model. The various conifold solutions arise by solving the algebraic constraints in diverse manners, and their previously known superpotentials, originally derived by other means, are given an explicit supergravity origin in this context. In this way, we extract the superpotential that generates the flow describing the baryonic branch of the Klebanov--Strassler theory, which was previously unknown. Furthermore, we obtain the superpotential for a supersymmetric solution describing D3 and wrapped D5-branes on the resolved conifold, that differs from the non-supersymmetric \hbox{one in \cite{PandoZayasTseytlin}.} 

The paper is organized as follows. In section~\ref{sec:setup} we present the theory under study, i.e.\ five-dimensional, $\mathcal N=4$ gauged supergravity. The main results regarding the general analysis of BPS domain walls are stated in section \ref{generalDWanalysis}, where we also present the analysis of the supersymmetry equations from the gravity multiplet. In sections \ref{dwIIBonSE} and \ref{dwT11} we apply our general conditions to the consistent truncations of type IIB string theory and construct their superpotentials and domain wall solutions. We conclude in section~\ref{Conclusions}. Some technical details on the $\mathcal N=4$ scalar manifold are given in appendix~\ref{appScalarManifoldGeom}, while appendix~\ref{AnalysisGaugino} contains the comprehensive analysis of the gaugino equation.


\section{$\mathcal N=4$ gauged supergravity in five dimensions}
\label{sec:setup}

The ungauged $\mathcal N=4$ supergravity theory in five dimensions was constructed in \cite{AwadaTownsend}, together with its SU(2)-gauged version. The generally gauged theory was given in \cite{DallHerrZag}, and further extended in \cite{SchonWeidner} by using the embedding tensor formalism (see e.g.\ \cite{Samtleben:2008pe} for a review).
In the following, we provide a compendium of the features that will be relevant. Since we will study domain wall solutions in which only the metric and the scalar fields acquire a non-trivial profile, it will be sufficient to discuss the couplings of the scalars and their appearance in the supersymmetry transformations of the fermions.

\subsection{Geometry of the scalar manifold}

The scalars of five-dimensional $\mathcal N=4$ supergravity coupled to $n$ vector multiplets define a $\sigma$-model whose target manifold is
\be
\mathcal M_{\rm scalars} \,=\, {\rm SO}(1,1)\times \frac{{\rm SO}(5,n)}{{\rm SO}(5)\times {\rm SO}(n)}\,.
\ee
The SO$(1,1)$ factor is parameterized by a real scalar $\sigma$, being part of the gravity multiplet, while the second factor is parameterized by real scalars $\phi^x$, $x=1,\ldots, 5n$, being part of the vector multiplets. The $\frac{{\rm SO}(5,n)}{{\rm SO}(5)\times {\rm SO}(n)}$ coset representative is $(\mathcal V_M{}^{\underline a},\mathcal V_M{}^{a})$, where $M=1,\ldots,5+n$ labels the standard representation of SO$(5,n)$, while $\underline a =1,\ldots,5 $ and $a =1,\ldots,n$ label the representations of SO(5) and SO$(n)$, respectively. So we have that the global SO$(5,n)$ acts from the left, while the local SO$(5)\times$SO$(n)$ acts from the right.

The coset manifold is equipped with an invariant metric and an ${\rm SO}(5)\times {\rm SO}(n)$ composite connection. These are encoded in the one-form $\mathcal V^{-1} d\mathcal V$, which takes values in the ${\rm so}(5,n)$ algebra and can be expressed as
\be
\mathcal V^{-1} d\mathcal V\,\,=\,\, 2\,\vielb^{a\ul b} t_{a\ul b} - \omega^{ab} t_{ab} + \omega^{\ul{ab}} t_{\ul{ab}}\,,
\ee
where we are splitting the $t_{MN}$ generators of ${\rm so}(5,n)$ into ${\rm so}(5)$ generators $t_{\ul{ab}}$, ${\rm so}(n)$ generators $t_{ab}$ and coset generators $t_{a\ul b}\,$. For definiteness, we work in the fundamental representation $(t_{MN})_P{}^Q= \delta^Q_{[M}\,\eta^{\phantom{Q}}_{N]P}$, where $\eta = {\rm diag}\{-----+\ldots+\}$ is the SO$(5,n)$ metric. 
In our expansion above,
\begin{equation}\label{vielbfromMCform}
\vielb^{a\ul b} \,\equiv\, \vielb_x^{a\ul a}\,d\phi^x\,=\, -\,(\mathcal V^{-1}  d\mathcal V)^{a\ul b}
\end{equation}
are the coset vielbeine, while
\be\label{connfromMCform}
\omega^{\ul{ab}} = (\mathcal V^{-1}  d\mathcal V)^{\ul{ab}}\,,\qquad \omega^{ab} =(\mathcal V^{-1}  d\mathcal V)^{ab}
\ee
are the SO(5) and SO$(n)$ pieces of the connection, respectively.
The vielbein defines an invariant metric $g_{xy}$ on the coset via 
\begin{equation}\label{relsvielb1}
\vielb_x^{a\underline a}\,\vielb_y^{a\underline a} \,=\, g_{xy}	 \,,
\end{equation}
and satisfies
\begin{equation}\label{relsvielb2}
\vielb_x^{a \underline a} \,\vielb^{x \,b \underline b} \,=\, \delta^{ab} \delta^{\underline{ab}} \,.
\end{equation}
%

The R-symmetry group of $\mathcal N=4$ supergravity in five dimensions is Spin(5)$\,\simeq\,$USp(4).
We denote by $i,j=1,...,4$ its spinorial representation, in which the fermion fields as well as the supersymmetry parameters transform.
In order to switch from spinor to vector representation, we use the Clifford map defined by the Cliff(5) Dirac matrices $(\Gamma^{\underline a})_i{}^j$. This acts on an antisymmetric $p$-tensor $T_{\ul{a_1\cdots a_p}}$ as
\be\label{defCliffordmap}
T_{\ul{a_1\cdots a_p}} \;\; \mapsto\;\; T_{ij}\, = \,T_{\ul{a_1\cdots a_p}} \Gamma^{\ul{a_1\cdots a_p}}_{\phantom{a_1\cdots a_p} ij}\,,
\ee
where $\Gamma^{\ul{a_1\cdots a_p}}$ is the antisymmetric product of $p$ $\Gamma$-matrices.
For instance, for the vielbein we have\footnote{We will have one exception to \eqref{defCliffordmap}: in order to match the expressions in \cite{AwadaTownsend}, we introduce a factor of 1/2 in the Clifford map for the coset representative, i.e.\ we define
\begin{equation}\nn
	{\cal V}_{M\,i}{}^j \,=\, \frac12 {\cal V}_M{}^{\underline a}\, \Gamma_{\underline a\,i}{}^{j}\,.
\end{equation}
}
\begin{equation}
	\vielb^{a\, ij} \,=\, \vielb^{a \underline a} \,\Gamma_{\underline a}{}^{ij}.
\end{equation}
Using the properties of the $\Gamma$-matrices summarized in appendix~\ref{appScalarManifoldGeom}, the relations \eqref{relsvielb1}, \eqref{relsvielb2} can equivalently be expressed as
\begin{equation}
	\vielb_x^{a\,ij}\vielb_{y\,ij}^a \,=\, 4\, g_{xy}\,, \qquad\qquad \vielb_x^{a\, ij} \vielb^{x\, b}{}_{kl} \,=\, \delta^{ab} \left(4\,\delta^{ij}_{kl}- \Omega^{ij} \Omega_{kl}\right).
\end{equation}
Here, $\Omega$ is the USp(4) invariant tensor, satisfying
\begin{equation}
\Omega_{ij} = - \Omega_{ji}\,,\qquad\quad \Omega_{ij}=\Omega^{ij}\,,\qquad\quad	\Omega_{ij} \Omega^{jk} = - \delta_i^k\,.
\end{equation}
This is used to raise and lower the indices, according to the NW-SE rule
\begin{equation}
	V^i = \Omega^{ij} V_j\,, \qquad V_i = V^j \Omega_{ji}\,.
\end{equation}

We will often omit the USp(4) spinorial indices $i,j$. It is understood that USp(4) matrices $S$, $T$ are multiplied according to the rule $S_i{}^k T_k{}^j$.
On the other hand, we will always explicitly display the vectorial indices $\ul{a},\,\ul{b}$. The $\ul a, \ul b$ (as well as the $a,b$) indices are always raised and lowered with the Kronecker delta.
Further useful relations satisfied by the quantities defined above are given in appendix~\ref{appScalarManifoldGeom}.

\subsection{Supersymmetry transformations}
\label{sec:susy_transformations}

Once the number of vector multiplets is fixed, the $\mathcal N=4$ gauged supergravity theory is fully characterized by the embedding tensor, which is a constant matrix specifying how the gauge group is embedded into the global symmetry group ${\rm SO}(1,1)\times {\rm SO}(5,n)$. 
The embedding tensor assigns charges to the different fields, and appears in various terms of the gauged supergravity action (gauge covariant derivatives, fermion mass terms, scalar potential) as well as in the supersymmetry transformations, where the extra terms due to the gauging are known as fermionic shifts.
As in \cite{SchonWeidner}, we denote its components by $f^{MNP}=f^{[MNP]}$, $\xi^{MN}=\xi^{[MN]}$ and $\xi^M$.

In the following we provide the supersymmetry transformations of the fermions, which will be our starting point for the analysis of supersymmetric domain walls.
The $\mathcal N=4$ supergravity fermions are the gravitini $\psi_{\mu\, i}\,$, the spin 1/2 fermions $\chi_i$, both sitting in the gravity multiplet, and the gaugini $\lambda_i^a$ in the vector multiplets. 
Note that they all carry a USp(4) index. Moreover, they are all symplectic-Majorana. We will restrict to the case in which all the one- and two-form fields in the theory, as well as the $\xi^M$ components of the embedding tensor, vanish. The latter assumption means that the SO(1,1) scalar $\sigma$ remains neutral. Then the variations given in \cite{SchonWeidner} read
\bea
\delta \psi_{\mu \,i} \!&=&\! D_\mu \epsilon_i -  \frac{i}{2}g\,  P_i{}^j \gamma_\mu \epsilon_j\,,\label{GravitinoVariation}\\[2mm]
\delta \chi_i \!&=&\! -\frac i2 \gamma^\mu\partial_\mu\sigma \, \epsilon_i + \frac{3}{2}g \, \partial_\sigma P_i{}^j \epsilon_j \,,\label{DilatinoVariation}\\[2mm]
	\delta \lambda_i^a \!&=&\! -\frac i2 \gamma^\mu\partial_\mu \phi^x \vielb^a_{x\,ij}  \epsilon^j  - g\,P^a{}_i{}^j \epsilon_j\,,\label{GauginoVariation}\qquad
\eea
where the $\mathcal N=4$ supersymmetry parameter $\epsilon_i$ is also a USp(4) symplectic-Majorana spinor, $\gamma^\mu$ are the Cliff$(1,4)$ Dirac matrices, $g$ is the gauge coupling constant (assumed positive), and we introduced
\be
\Sigma = e^{\sigma/\sqrt3}\,.
\ee 
The derivative $D_\mu$ is covariant both with respect to the SO$(1,4)$ Lorentz and the USp(4) R-symmetry transformations on the spinor $\epsilon_i$. Moreover, we defined the gravitino shift matrix
\begin{equation}\label{defPij}
P_{ij} \,=\, P^{\underline{ab}}\,\Gamma_{\underline{ab}}{}_{\,ij}\,, 
\end{equation} 
with
\begin{equation}\label{defPab}
P^{\underline{ab}} \,:=\,  -\frac{1}{6\sqrt 2}\Sigma^2 \xi^{\underline{ab}} + \frac{1}{36}\Sigma^{-1}\epsilon^{\underline{abcde}}f^{\underline{cde}}  \,,
\end{equation}
where we are using the ``dressed'' components of the embedding tensor
\begin{equation}
\xi^{\underline{ab}} := \xi^{MN}\mathcal V_M{}^{\underline a} \mathcal V_N{}^{\underline b} \;,\qquad\quad
f^{\underline{abc}} := f^{MNP}\mathcal V_M{}^{\underline a} \mathcal V_N{}^{\underline b} \mathcal V_P{}^{\underline c}\,.
\end{equation}
The USp(4) matrix $P_{ij}$ is going to be our ``matrix superpotential'', out of which we will extract the actual superpotential generating the flow equations.
We have also introduced the gaugino shift matrix
\be\label{gauginoshift}
P^a_{ij} \,:=\, \frac{1}{\sqrt{2}}\,\Sigma^2\, \xi^{MN}\mathcal{V}_{M}{}^a\mathcal{V}_{N\,ij} \,+\, \Sigma^{-1} f^{MNP}\,\mathcal{V}_M{}^a\,\mathcal{V}_{N\,i}{}^k\,\mathcal{V}_{P\,kj}\,.
\ee
This can only in part be written as a derivative of the gravitino shift (see appendix~\ref{AnalysisGaugino}).

\section{Supersymmetric domain walls}\label{generalDWanalysis}

In this section, we come to study the conditions for supersymmetric domain walls.

\subsection{General strategy and preliminary decomposition}\label{sec:decomposition}

We specialize to a (flat) domain wall ansatz, namely we take a spacetime metric of the form~\eqref{dwmetric} and we assume that all scalars as well as the supersymmetry parameter $\epsilon_i$ just depend on the fifth, radial coordinate $r$. We also switch off all one- and two-forms in the theory. 

The supersymmetry conditions follow from setting to zero the fermion variations given in the previous section. In detail, the $\mathbb{R}^{1,3}$ components of the gravitino variation \eqref{GravitinoVariation} yield the condition
\be
 -A'\gamma_5 \epsilon_i +  g \,i\, P_{i}{}^j \epsilon_j\;=\; 0\,,\label{GravitinoEqNonRadial} 
\ee
while the radial component imposes
\be
\epsilon_i' + \phi^{x\,\prime} \omega_{x\, i}{}^j \epsilon_j - \frac{i}{2} g\, P_{i}{}^j \gamma_5 \epsilon_j\;=\; 0\,,\label{GravitinoEqRadial}
\ee
where we recall that $\omega_x{}_i{}^j$ is the SO(5) connection on the scalar manifold.
The variation~\eqref{DilatinoVariation} of the other fermion in the gravity multiplet yields
\be
 \sigma' \,\gamma_5 \epsilon_i + 3ig \, \partial_\sigma P_{i}{}^j \epsilon_j\;=\; 0\,, \label{DilatinoEq}
\ee
and finally the gaugino variation~\eqref{GauginoVariation} gives
\be\label{GauginoEq}
i \phi^{x\prime}  \vielb^a_{x\,ij}\gamma_5\epsilon^j  + 2g\, P^a{}_i{}^j\epsilon_j\;=\; 0\,.
\ee

A special role in our analysis will be played by the gravitino shift matrix  \eqref{defPij}, \eqref{defPab}. As we will show, fixing it generically reduces the R-symmetry of the solution from USp(4) to U(1), namely it produces 1/4 BPS configurations. Furthermore, its eigenvalues correspond to the possible superpotentials generating the flow equations. In the following, we split the gravitino matrix in a way that is most suitable for making these features manifest.

Let us first describe how a ${\rm USp}(4)\to {\rm SU}(2)$ breaking is defined.
The gravitino shift matrix can be used to construct an SO(5) vector 
\be\label{Xdefinition}
\widetilde{\vct}^{\underline{a}} \,=\, \epsilon^{\underline{abcde}} P_{\underline{bc}} P_{\underline{de}}\,,
\ee
whose norm is
\be
|\widetilde X| \,\equiv\, \sqrt{\widetilde{\vct}^{\ul a}\widetilde{\vct}_{\ul a}} \,=\, \sqrt{8\,(P^{\underline{ab}}P_{\underline{ab}})^2-16\,P^{\underline{ab}}P_{\underline{bc}}P^{\underline{cd}}P_{\underline{da}} }\;.
\ee
Away from the zero locus, we can introduce a normalized vector
\be
\vct \,=\, \widetilde{\vct}\,/\,|\widetilde X|\,.
\ee
which pointwise on the scalar manifold specifies a preferred direction in $\mathbb R^5$ and therefore an SO(4) subgroup of the R-symmetry group SO(5).
On the spinors, this defines a reduction USp(4) $\to {\rm SU}(2)_+\!\times {\rm SU}(2)_-$, where the plus and minus refer to the chiralities defined by $\vct_i{}^j = \vct_{\ul{a}}\Gamma^{\ul{a}}{}_i{}^j$. The projectors over these SU(2)$_\pm$ are
\be
\Pi_\pm{}_i{}^j \,=\, \frac{1}{2}\left( \delta_i{}^j \pm \vct_i{}^j\right)
\ee
(note that the projector condition $\Pi_\pm^2 = \Pi_\pm$ is satisfied since $\vct_i{}^k\vct_k{}^j = \delta_i{}^j$). As we will see, on supersymmetric domain walls either one of these projectors acts on the Usp(4) supersymmetry parameter, constraining it to be an SU(2) spinor.

The matrix $P$ actually constrains the supersymmetry parameter even further.
We note that $P^{\ul{ab}}$ lives in the so(4) identified by $\vct$, since it is satisfied that
\be\label{APvanishes}
\vct_{\ul a}P^{\ul{ab}} = 0\,,
\ee 
which follows from the definition of $\vct$ together with the identity
$\epsilon^{\underline{abcde}} T_{\underline{ab}} T_{\underline{cd}} T_{\underline{ef}} = 0\,,$
holding for any antisymmetric tensor $T_{\underline{ab}}$ in five dimensions. Note that \eqref{APvanishes} can also be written as
\be
\vct\, P \,=\, P\vct\,.
\ee
Hence the $P$ matrix can be decomposed as the sum of two matrices in ${\rm so}(4)= {\rm su}(2)_+\oplus {\rm su}(2)_-$:
\be\label{PintoPplusPminus}
P = P_+ + P_-\,,
\ee
with the su(2)$_\pm$ components being
\be
P_\pm \,:=\, P \, \Pi_\pm \,=\, \Pi_\pm P \,.
\ee
Squaring $P= P_{\underline{ab}} \Gamma^{\ul{ab}}$, we find\footnote{The fact that the square of the gravitino shift matrix $P$ is not just proportional to the identity is a crucial difference with respect to the $\mathcal N=2$ analysis of \cite{5dN=2domainwalls}. Of course, this is because the former is a USp(4) matrix while the latter is an SU(2) matrix. This also motivates why in the $\mathcal N=4$ case we generically obtain 1/4 BPS domain walls while in the $\mathcal N=2$ case 1/2 BPS configurations are found.}
\be\label{Psquare}
P^2 \,=\, - 2 P_{\underline{ab}} P^{\underline{ab}} \, 1_4 + \widetilde\vct\,.
\ee
Noting that $P_+ P_- =0$, we deduce 
\be\label{Ppmsquare}
P_\pm^2 \, = \, - W_\pm^2\, \Pi_\pm \,,
\ee
where we defined the real functions of the scalar fields
\bea
W_{\pm} &=& \sqrt{2\,P^{\underline{ab}}P_{\underline{ab}}\mp |\widetilde X|} \nn\\[2mm]
&=& \sqrt{2\,P^{\underline{ab}}P_{\underline{ab}}\mp \sqrt{8\,(P^{\underline{ab}}P_{\underline{ab}})^2-16\,P^{\underline{ab}}P_{\underline{bc}}P^{\underline{cd}}P_{\underline{da}} }} \label{eq:ExprW}\;.
\eea
One can see that $+W_\pm$ and $- W_\pm$ coincide with the four real eigenvalues of the hermitian matrix $i P$. Either $W_+$ or $W_-$ will play the role of the superpotential generating the flow equations.
For $W_\pm\neq 0$, we can introduce normalized matrices
\be\label{defQpm}
Q_\pm \,=\,\frac{i}{W_\pm} P_\pm 
\ee
which satisfy
\be\label{Qpmsquare}
Q_\pm^2 \,=\, \Pi_\pm\,
\ee
as well as
\begin{equation}\label{projectedQ}
\vct Q_{\pm} \,=\, Q_{\pm} \vct \,=\, \pm Q_{\pm}\,.
\end{equation}
So we have 
\be\label{PisWQplusWQ}
P = -i\, W_+ Q_+ -i\, W_- Q_-\,.
\ee
As we will discuss below, the $Q$ matrix will be involved in an algebraic constraint acting on the USp(4) spinor parameter, breaking the R-symmetry of the solution down to U(1).

It will be useful to also introduce generators $(L^r_\pm)_i{}^j$, $r=1,2,3$, for the su(2)$_\pm$ subalgebras of so(5) identified by $\vct$. These verify\footnote{
An explicit realization of these su(2)$_\pm$ generators is the following. First we note that the unit SO(5) vector $\vct$ can be put along e.g.\ the fifth direction by exploiting the local SO(5) symmetry transformations which act on the ${\rm SO}(5,n)/({\rm SO}(5)\times{\rm SO}(n))$ coset representative from the right. This could in principle be done only at a fixed value of the SO(1,1) scalar $\sigma$, because a priori $\vct$ depends on it. However, in our analysis we will find that on supersymmetric domain walls, $\vct$ has to be independent of $\sigma$, see eq.~\eqref{dsigmavanishes} below. Hence we can take $\vct_{\underline a} = \delta^5_{\underline a}\,$, or in spinorial indices $\vct_i{}^j =\Gamma_{5}{}_{\,i}{}^j\,$.
Then the su(2)$_\pm$ generators can be taken as:
\begin{eqnarray}
L_\pm^1 = \frac{1}{2} \left( \pm\Gamma_{12
} - \Gamma_{34} \right)\,,\qquad L_\pm^2 = \frac{1}{2} \left( \pm\Gamma_{13} + \Gamma_{24} \right)\,,\qquad 
L_\pm^3 = \frac{1}{2} \left( \pm\Gamma_{14} - \Gamma_{23}\right)\,.
\end{eqnarray}
\label{footnoteSO5gauge}}
\be
\label{eq:XL} \vct \,L^r_\pm \,=\, L^r_\pm \vct \,= \,\pm L^r_\pm\,,
\ee
\be
\label{eq:LL} L^r_\pm L^s_\pm \,=\, -\,\delta^{rs} \Pi_\pm  + \, \epsilon^{rst} L^t_\pm\,,
\ee
\be\label{eq:LplusLminus} 
L^r_+ L^s_- \,=\, 0\,,
\ee
together with the completeness relation
\be\label{completenessL}
L^r_\pm{}_{ij} L^r_\pm{}^{kl} \,=\, 2 \,\Pi_\pm{}_i{}^{(k} \, \Pi_\pm{}_j{}^{l)}\,.
\ee
So our $P$ can be written as
\be\label{expansionPasWQL}
P = P^r_+ L^r_+ + P^r_- L^r_- \, = -iW_+ Q^r_+ L^r_+ -iW_- Q^r_- L^r_-\,.
\ee

Having split the gravitino shift matrix in a suitable way and having introduced the main characters playing a role in our analysis, we can now present our results.

\subsection{The necessary and sufficient conditions}\label{NecSufCond}

Our main result in this first part of the paper is that the equations \eqref{GravitinoEqNonRadial}--\eqref{GauginoEq} for a supersymmetric domain wall in $\mathcal N=4$ gauged supergravity are equivalent to a system of first-order flow equations for the scalar fields, generated by the superpotential $W_+$, or $W_-$, introduced in eq.~\eqref{eq:ExprW} above, and supplemented by a set of constraints. 
Specifically, if $W_+$ is chosen, we find the conditions
\begin{eqnarray}
	 A'  &=& \pm\, g\,W_+\,, \label{metricflow}\\ [2mm]
	 \sigma' &=& \mp\, 3g\,\partial_\sigma W_+\,, \label{sigmaflow}\\ [2mm]
\phi^{x\,\prime} &=& \mp\, 3g\,g^{xy}\,\partial_y W_+\,,	 \label{FloweqFromGaugino_MainText}
\end{eqnarray}
together with
\begin{equation}\label{dsigmavanishes}
	\partial_\sigma \vct \,=\, 0 \,=\, \partial_\sigma Q^r_{+}\,, 
\end{equation}
\be\label{covderAiszero}
\phi^{x\,\prime} D_x \vct \,=\, 0\,=\, \phi^{x\,\prime} D_x Q^r_+\,,
\ee
(which can equivalently be written as $\partial_\sigma Q_+ = 0$ and $\phi^{x\,\prime}D_x Q_+ =0$, respectively)
and
\bea
&&\xi^{a\ul{a}}\vct_{\ul a} \,=\, 0\,,\label{condition_xi}\\ [2mm]
&&f^{a\ul{ab}} L^r_{+ \ul{ab}}\left(\delta^{rs} + Q_+^r Q_+^s \right) \,=\, 0\,.\label{condition_f}
\eea
In the last two relations we used the dressed components of the embedding tensor
\be
\xi^{a\ul{a}}:= \xi^{MN}\mathcal V_M{}^{a}\mathcal V_N{}^{\ul{a}}\,,\qquad f^{a\ul{ab}} := f^{MNP}\,\mathcal{V}_M{}^a\,\mathcal{V}_N{}^{\ul{a}}\,\mathcal{V}_P{}^{\ul{b}}\,.
\ee
Moreover, the $\mathcal N=4$ supersymmetry parameters are constrained by
\be\label{projectionspinors}
Q_+{}_i{}^j\epsilon_j \,=\, \pm\, \gamma_5\epsilon_i\,,
\ee
which means that the domain wall preserves four out of sixteen supercharges, i.e. it preserves $\mathcal N=1$ supersymmetry. This can be seen by first noting that by eq.~\eqref{Qpmsquare} our constraint implies $\vct_i{}^j \epsilon_j = \epsilon_i$. This means that the supersymmetry parameter $\epsilon_i$, which generically is a USp(4) symplectic-Majorana spinor carrying sixteen real degrees of freedom, is actually restricted to be an SU(2) symplectic-Majorana spinor, carrying eight real degrees of freedom. Then one observes that on such spinors $Q_+$ squares to the identity and therefore \eqref{projectionspinors} defines a projector halving the number of degrees of freedom, so that we are left with four independent supercharges.

As we will discuss below, these supersymmetry conditions imply the equations of motion of $\mathcal N=4$ supergravity and we therefore have a full solution of the five-dimensional theory.

A few comments on the constraints. The first condition \eqref{dsigmavanishes} tells us that the choice of ${\rm so}(4)\subset {\rm so}(5)$ does not depend on~$\sigma$, while the second one constrains the projector related to the domain wall direction.\footnote{The latter is analogous to the condition $\partial_x Q^r=0$ found in the $\mathcal N=2$ analysis of \cite{5dN=2domainwalls}, while the former was not present in that case.} Conditions \eqref{covderAiszero} tell that $\vct$ and $Q$ are constant with respect to the SO(5) covariant derivative evaluated along the flow.
Finally, for each value of the index $a$, namely for each vector multiplet, eq.~\eqref{condition_xi} states that the so(5) vector $\xi^{a\ul{a}}$ is orthogonal to $\vct_{\ul a}$, while \eqref{condition_f} tells that the su(2)$_+$ piece of the so(5) algebra valued matrix $f^{a\ul{ab}}$ has to be along $Q\,$.

If $W_-$ is chosen instead of $W_+$, one has the same conditions \eqref{metricflow}--\eqref{projectionspinors}, with the only change that $W_+, Q_+, L^r_+$ are everywhere replaced by $W_-, Q_-, L^r_-$, respectively. Notice that in this case $\vct_i{}^j \epsilon_j = -\epsilon_i$.

The flow equation~\eqref{metricflow} for the warp factor, the one \eqref{sigmaflow} for the scalar $\sigma$, and the constraints \eqref{dsigmavanishes}, \eqref{covderAiszero}, \eqref{projectionspinors}, arise from the supersymmetry variation of the fermions in the $\mathcal N=4$ gravity multiplet. This is proved in the next section. The other conditions, namely the flow equations for the remaining scalar fields~\eqref{FloweqFromGaugino_MainText} and constraints~\eqref{condition_xi}, \eqref{condition_f}, come from the variation of the gaugini in the vector multiplets. The complete proof of this statement is very non-trivial, and represents a major piece of our analysis; however since it involves many technical steps we present it in appendix~\ref{AnalysisGaugino}.
There, we also show that the second in~\eqref{covderAiszero} is implied by the gaugino equation (see eq.~\eqref{DxQ}), so it does not need to be checked independently.

Finally, we remark that the results above assume $|\widetilde X| \neq 0$. In the special case $|\widetilde X|=0$, which implies $\widetilde\vct =0$, the projection $\Pi_{\pm}$ on ${\rm SU(2)}_\pm$ cannot be defined and the analysis proceeds in a slightly different way. Though in this case we have not performed a complete study of the constraints arising from the gaugino equation, one can see that the first-order flow equations for the warp factor and the scalar fields still hold, but at the place of $W_\pm$ there is just one possible superpotential, $W = \sqrt {2P_{\ul{ab}}P^{\ul{ab}}}$, and the gravitino shift matrix can be written as $P = -iW Q$, with $Q^2 = 1_4$. The only constraint on the supersymmetry parameter from the gravitino equation is $Q_i{}^j\epsilon_j = \gamma_5 \epsilon_i$, which is a projection halving the amount of supersymmetry. So in this $|\widetilde X| = 0$ case we expect 1/2 BPS domain walls, preserving eight Poincar\'e supercharges.

\subsection{Analysis of the equations from the gravity multiplet}\label{AnalysisGravityMult}

In this section, we prove that eqs.~\eqref{GravitinoEqNonRadial}--\eqref{DilatinoEq}, ensuring that the variation of the fermions in the gravity multiplet vanishes, are equivalent to eqs.~\eqref{metricflow}, \eqref{sigmaflow}, \eqref{dsigmavanishes}, \eqref{covderAiszero} and \eqref{projectionspinors} when $W_+$ is chosen. The proof for $W_-$ works in the same way.

We start from the gravitino equation \eqref{GravitinoEqNonRadial}, i.e.\begin{equation}\label{eq:gravitinoeq}
g\,i\, P_{i}{}^j \epsilon_j \,=\,	A'\gamma_5 \epsilon_i\,.
\end{equation}
Applying $P$ to this equation and using \eqref{Psquare}, we obtain
\be
- g^2 \widetilde\vct_i{}^j  \epsilon_j = \left[\left(A'\right)^2 -2 g^2 P_{\underline{ab}} P^{\underline{ab}} \right] \epsilon_i \,.
\ee
Applying $\widetilde \vct$ to this equation, one immediately sees that its solution is the flow equation \eqref{metricflow} for the warp factor $A\,$:
\be\label{warpflow}
A' \,= \, \pm\, g\,W_+   \,,
\ee
with $W_+$ as in \eqref{eq:ExprW}, and with $\epsilon_i$ satisfying
\be\label{Aepsiseps}
\vct{}_i{}^j \epsilon_j \,=\, \epsilon_i\,.
\ee
The latter condition constrains the USp(4) spinor $\epsilon_i$ to be a spinor of ${\rm SU}(2)_+$. Note that if $|\widetilde X|=0$ this condition does not hold. In the following, we will assume $|\widetilde X|\neq 0$. Also note that we could equally well have taken $W_-$ in \eqref{warpflow}, and consequently $\vct{}_i{}^j \epsilon_j \,=\, -\epsilon_i$, implying that $\epsilon_i$ is a spinor of SU(2)$_-$. In the following we keep on discussing the case in which $W_+$ is chosen. The argument for the opposite choice is completely analogous and just requires to switch a few signs. 

With these assumptions, recalling \eqref{PintoPplusPminus}, \eqref{defQpm}, and noting that \eqref{Aepsiseps} implies \hbox{$Q_-{}_i{}^j\epsilon_j = 0$}, in the locus where $W_+ \neq 0$ the original gravitino equation \eqref{eq:gravitinoeq} becomes just constraint \eqref{projectionspinors}:
\be\label{projectorQ}
Q_+{}_i{}^j \epsilon_j \, = \, \pm\,\gamma_5\epsilon_i\,.
\ee
From \eqref{Qpmsquare} we see that this acts as a projector on the SU(2)$_+$ spinor, halving its number of degrees of freedom.
Also note that this condition actually implies \eqref{Aepsiseps}.

\medskip

We next consider equation~\eqref{DilatinoEq} arising from the transformation of the fermions $\chi_i$ in the gravity multiplet, namely
\begin{equation}\label{eq:EigvProbldsigmaP}
3 i\,g \, \partial_\sigma P_{i}{}^j \epsilon_j \,=\,  - \sigma' \,\gamma_5 \epsilon_i \,.
\end{equation}
Recalling \eqref{PisWQplusWQ} and the algebraic constraints on the spinor obtained above, this can be rewritten as
\begin{equation}\label{DilatinoEqWithQ}
3g\,\partial_\sigma W_+Q_+{}_i{}^j\epsilon_j +3g\, W_+\partial_\sigma Q_+{}_i{}^j\epsilon_j + 3gW_-\partial_\sigma Q_-{}_i{}^j \epsilon_j \,=\, \mp\,\sigma'\,Q_+{}_i{}^j\epsilon_j\,.
\end{equation}
Applying either $(W_+Q_++W_-Q_-)$ or $W_+\gamma_5$, using again the constraints on the spinor, adding the two resulting expressions and noting that $W_+^2-W_-^2=2|\widetilde X|$, we arrive at 
\begin{equation}\label{ElaboratingSigmaEq}
6g\,W_+\partial_\sigma W_+\epsilon_i +3g|\widetilde X|\,\partial_\sigma\vct_i{}^j\epsilon_j \,=\, \mp\,2W_+\sigma'\epsilon_i\,.
\end{equation}
We also observe that $\vct^2 = 1_4$ implies $\partial_\sigma \vct \vct = - \vct\partial_\sigma \vct $, which applied to $\epsilon$ gives
\be
\Pi_+\partial_\sigma \vct\,\epsilon \,=\, \partial_\sigma \vct \, \Pi_- \epsilon \,=\, 0\,.
\ee
Hence projecting eq.~\eqref{ElaboratingSigmaEq} with $\Pi_+$, for $W_+\neq 0$ we obtain the flow equation \eqref{sigmaflow} for the scalar~$\sigma$,
\begin{equation}
\sigma' \,=\, \mp\,3g\,\partial_\sigma W_+\,.
\end{equation}
Then eq.~\eqref{ElaboratingSigmaEq} becomes $\partial_\sigma \vct_i{}^j\epsilon_j = 0$, which is equivalent to
\begin{equation}
	\label{d_sigmaA}
\partial_\sigma \vct \,=\,0\,.
\end{equation}
This means that the choice of so(4) inside so(5) does not depend on $\sigma$.
Taking the derivative of \eqref{projectedQ}, it follows that $\partial_\sigma Q_+$ and $\partial_\sigma Q_-$ are in the su(2)$_+$ and su(2)$_-$ subspaces respectively:
\begin{equation}
\vct \partial_\sigma Q_{\pm}  \, =\, \partial_\sigma Q_{\pm} \vct \, =\, \pm \partial_\sigma Q_{\pm}\, .
\end{equation}
Consequently we deduce that
\be
\partial_\sigma Q_{-}\epsilon \,=\,  \partial_\sigma Q_{-}\Pi_-\epsilon  \,=\,  0\,.
\ee
Using the conditions just derived, our original equation \eqref{DilatinoEqWithQ} reduces to 
\begin{equation}\label{dilatinoeqreduced}
\partial_\sigma Q_{+}{}_i{}^j\epsilon_j=0\,.
\end{equation}
Since the choice of so(4) inside so(5) does not depend on $\sigma$, our su(2)$_\pm$ generators $L^r_\pm$ do not depend on $\sigma$ either, so we can write $\partial_\sigma Q_+ = \partial_\sigma Q^r_+ L^r_+$. Then, applying twice condition \eqref{dilatinoeqreduced} and using property \eqref{eq:LL} of the su(2) generators, we find
\be\label{stepstodQr=0}
\partial_\sigma Q^r_+\partial_\sigma Q^s_+ L^r_+L^s_+\epsilon =0 \quad\Rightarrow\quad \partial_\sigma Q^r_+\partial_\sigma Q^r_+ =0\quad\Rightarrow\quad \partial_\sigma Q^r_+ =0\,.
\ee
We have thus also obtained both constraints \eqref{dsigmavanishes}. Note that they can equivalently be stated as $\partial_\sigma Q_+ = 0$.

\medskip

Finally, let us consider eq.~\eqref{GravitinoEqRadial} for the radial flow of the spinor,
\be
\epsilon_i' + \phi^{x\prime} \omega_{x\, i}{}^j \epsilon_j - \frac{i}{2} g\, P_{i}{}^j \gamma_5 \epsilon_j\;=\; 0\,,
\ee
where we recall that $\omega_i{}^j$ is the USp(4) connection.
It is not hard to see that compatibility with the algebraic condition~\eqref{projectionspinors} requires
\be\label{D_xQvanishesOnFlow}
\phi^{x\,\prime} D_x Q_+{}_i{}^j \epsilon_j \, = \, 0\,.
\ee
Note that acting with $\phi^{x\,\prime}D_x$ on~\eqref{Qpmsquare} and using~\eqref{projectionspinors},~\eqref{D_xQvanishesOnFlow}, we obtain $\phi^{x\,\prime} D_x \vct_i{}^j \epsilon_j =0$, which is equivalent to
\be\label{D_xVctVanishesOnFlow}
\phi^{x\,\prime} D_x \vct \,=\, 0\,,
\ee
namely the SO(5) vector $\vct$ has to be covariantly constant along the flow. It follows that the  so(4) generators $L^r_\pm$ can be chosen to satisfy $\phi^{x\,\prime}D_x L^r_\pm{} = 0$.
Then by the same argument used in \eqref{stepstodQr=0}, eq.~\eqref{D_xQvanishesOnFlow} reduces to 
\be\label{D_xQplusvanishesOnFlow}
\phi^{x\,\prime}D_x Q^r_+ = 0\,.
\ee 
We have thus also obtained constraints \eqref{covderAiszero}. They are equivalent to $\phi^{x\,'}D_xQ_+=0$.

Taking \eqref{D_xQvanishesOnFlow} into account, and using the flow equation~\eqref{metricflow}, the spinor equation becomes
\be\label{radialeqcompatible}
\left( e^{-A/2}\epsilon_i\right)' + \phi^{x\,\prime} \omega_x{}_i{}^j \left( e^{-A/2}\epsilon_j\right) \,=\, 0\,.
\ee
By performing a local SO(5) transformation, we could set to zero the component of the SO(5) connection along the flow, so that the covariant derivatives in~\eqref{D_xVctVanishesOnFlow} and~\eqref{D_xQplusvanishesOnFlow} become partial derivatives, and the solution for the spinor is
\be
\epsilon_i \, = \, e^{A/2}\, \tilde \epsilon_i\,,
\ee
where $\tilde \epsilon_i$ is a constant spinor satisfying the algebraic constraint~\eqref{projectionspinors}.\footnote{If part of the gauge freedom has already been fixed so that the framework of footnote~\ref{footnoteSO5gauge} is achieved and $\vct$ is made constant on the scalar manifold, then eq.~\eqref{D_xVctVanishesOnFlow} becomes a {\it requirement} that the piece of the SO(5) connection taking values in the orthogonal complement of so(4) in so(5) vanishes. The components of the connection appearing in~\eqref{D_xQplusvanishesOnFlow} and~\eqref{radialeqcompatible} can still be set to zero exploiting the residual gauge freedom.}

\medskip

This concludes our analysis of the supersymmetry conditions arising from the gravity multiplet. As already mentioned, the full analysis of the gaugino equation is technically more involved, so that we relegated it to appendix~\ref{AnalysisGaugino}.

\subsection{The domain wall effective action}

Plugging in the domain wall ansatz, the five-dimensional $\mathcal N=4$ supergravity action (for which we refer to \cite{DallHerrZag, SchonWeidner}) reduces to the one-dimensional effective action
\be
S \,=\, \frac{1}{2}\int_{-\infty}^{+\infty} dr \,e^{4A}\left[ 12(A')^2 - (\sigma')^2 - g_{xy}\phi^{x\,\prime}\phi^{y\,\prime} -2V \right]
 -4 \left[e^{4A}A'\right]_{-\infty}^{+\infty}
\ee
whose equations of motion are the same as the Einstein and the scalar equations of five-dimensional supergravity, evaluated in the domain wall ansatz.

Starting from the general formula for the $\mathcal N=4$ scalar potential $V$ written as an algebraic sum of squares of the fermionic shifts (see \cite{SchonWeidner}), with a long computation we found that this can be rewritten as the following algebraic sum of quadratic terms
\bea
g^{-2} \,V \!&=&\!\frac92\, g^{xy}\partial_x W\partial_yW +\frac92\,\left(\partial_\sigma W\right)^2-6\,W^2-18\,W^2\, \partial_\sigma Q_{\underline{ab}} \partial_\sigma Q^{\underline{ab}}  -9\,|\widetilde X|\left(\partial_\sigma \vct_{\underline{a}}\right)^2\nonumber\\[3mm]
\!&&\!+\frac{1}{4}\,\Sigma^4\left(\xi^{a\underline{a}}\vct_{\underline{a}}\right)^2+2\, \Sigma^{-2}\,{\rm Tr}\left[f^{a\underline{ab}}L^r{}_{\underline{ab}}\left(\delta^{rs}+Q^rQ^s\right)\right]^2,
\eea
where $W,Q,L^r$ can be either $W_+,Q_+,L^r_+\,$, or $W_-,Q_-,L^r_-$.
It follows that our effective action can be recast in a BPS form as
\bea
S \!\!&=&\!\! \frac{1}{2}\int_{-\infty}^{+\infty} \!\!dr \,e^{4A} \Big\{ 12\left(A'\mp g\,W\right)^2 - \left(\sigma'\pm 3g\,\partial_\sigma W\right)^2 - g_{xy}\left(\phi^{x\,\prime} \pm 3g\,\partial^x W\right)\left(\phi^{y\,\prime}\pm 3g\,\partial^y W\right) \nn\\ [3mm]
\!\!&+&\!\!\!\! 36W^2 \partial_\sigma Q_{\underline{ab}} \partial_\sigma Q^{\underline{ab}} +18|\widetilde X|\left(\partial_\sigma \vct_{\underline{a}}\right)^2 -\frac{1}{2}\Sigma^4\left(\xi^{a\underline{a}}\vct_{\underline{a}}\right)^2 - 4\Sigma^{-2}{\rm Tr}\!\left[f^{a\underline{ab}}L^r{}_{\underline{ab}}\left(\delta^{rs}+Q^rQ^s\right)\right]^2\! \Big\} \nn \\[3mm]
\!\!&+&\!\!\!\! \left[e^{4A}(\pm 3 g\, W -4A')\right]_{-\infty}^{+\infty}\,.
\label{actionBPSform}
\eea
Since it is an algebraic sum of squares of the supersymmetry conditions summarized in section~\ref{NecSufCond}, clearly it is extremized by them.
It follows that the Einstein and scalar equations of motion of five-dimensional $\mathcal N=4$ supergravity are solved. In order to have a full solution, in remains to consider the equations of motion for the vector and two-form fields in the theory. The equations for the two-forms are trivially satisfied in the domain wall background. As for the equations for the vector fields, the only non-trivial contribution comes from the scalar kinetic term $g_{xy}D_\mu\phi^{x} D^\mu\phi^{y}$, where the gauge covariant derivative is $D_\mu\phi^x = \partial_\mu\phi^x + \Sigma^{-2}K^x A_\mu^0 + \Sigma K^x_M A_\mu^M$. Here, $A^0$ and $A^M$ are the $\mathcal N=4$ vector fields, while the $\Sigma^{-2} K$ and $\Sigma K_M$ are the Killing vectors on the scalar manifold which generate the isometries being gauged (see appendix~\ref{AnalysisGaugino} for their explicit expression). The vector equations then read
\be
\phi^{x\,\prime} K_x \,=\, 0\,,\qquad \phi^{x\,\prime} K^M_x \,=\, 0\,,
\ee
which, using the flow equations, coincide with the requirement of gauge invariance of the superpotential $W$. Hence we conclude that for a gauge-invariant superpotential also the vector field equations are satisfied and our conditions yield a solution to $\mathcal N=4$ supergravity.

\newpage

\section{$\!\!$Deforming SCFT's on D3-branes at Calabi-Yau cones}\label{dwIIBonSE}

We now apply the results of our general analysis, in order to study supersymmetric domain wall solutions, to two specific $\mathcal N=4$ supergravity models, which arise as consistent truncations of type IIB supergravity and are relevant for the gauge/gravity correspondence. The first, to be studied in this section, can be defined as a dimensional reduction on any five-dimensional Sasaki--Einstein manifold, while the second, which is an extension of the former, is based on the $T^{1,1}$ coset space, and will be discussed in the next section.

In~\cite{IIBonSE} (see also \cite{GauntlettVarelaIIBonSE,LiuSzepietowskiZhaoIIBonSE,SkenderisTaylorTsimpis,Liu:2010pq,Bah:2010cu}), a universal consistent truncation of type IIB supergravity on five-dimensional manifolds admitting a Sasaki--Einstein (SE) structure was constructed, leading to gauged $\mathcal N=4$ supergravity in five dimensions. This model admits the standard AdS$_5$ solution of type IIB on a Sasaki--Einstein manifold, and is thus suitable to describe universal deformations of the superconformal field theory on D3-branes probing the tip singularity of the Calabi--Yau cone over the Sasaki--Einstein space. It comprises two $\mathcal N=4$ vector multiplets and hence eleven scalar fields, dubbed $U,V,\phi, C_0, a,b^J,c^J$ (real) and $b,c$ (complex) in~\cite{IIBonSE}.\footnote{$\,b,c$ were actually denoted $b^\Omega$, $c^\Omega$ in \cite{IIBonSE}.} Here, $C_0$ and $\phi$ are the axion and dilaton of type IIB supergravity, while the other scalars enter in the ten-dimensional metric as
\begin{equation}\label{eq:10dmetric}
ds^2_{10} \,=\, e^{-\frac{2}{3}(4U+V) + 2A} ds^2(\mathbb{R}^{1,3}) + e^{-\frac{2}{3}(4U+V)} dr^2 \,+\,e^{2U}ds^2(\mathcal B_{\rm KE})\,+\,e^{2V}\eta^2\,,
\end{equation}
in the type IIB two-form potentials as
\begin{equation}
	B \,=\, b^{J} J+ {\rm Re}(b \,\Omega)\,,\qquad \quad	C_2 \,=\, c^{J} J+ {\rm Re}(c \,\Omega)\,,
\end{equation}
and in the five-form field strength as
\be
F_5 = (1+*_{10})\Big\{ \big[3\,{\rm Im}(b \overline{c }) + k\big] J\wedge J\wedge\eta +  \frac{1}{2}\big[ 2 da +  b^J  d c^J -c^Jdb^J + {\rm Re}\!\left(b  \,d\overline{c } - c \, d\overline{b }\right) \big]\wedge J\wedge J \Big\}.
\ee
Here, $\mathcal B_{\rm KE}$ is the four-dimensional K\"ahler--Einstein base of the Sasaki--Einstein manifold, $J$ is its K\"ahler form, $\eta$ is the contact one-form dual to the Reeb vector, satisfying $d\eta = 2J$, and $\Omega$ is the complex two-form which satisfies $d\Omega = 3i\, (\eta \wedge\Omega)$. The constant $k$ parameterizes the five-form flux. Note that in the metric we already assumed a five-dimensional domain wall ansatz.

The scalar in the $\mathcal N=4$ gravity multiplet is identified with
$
\sigma = -\frac{2}{\sqrt 3}(U+V)\,,
$
while the others parameterize the $\frac{{\rm SO}(5,2)}{{\rm SO}(5)\times {\rm SO}(2)}$ coset space, see \cite{IIBonSE} for details.
In the parameterization chosen there, the non-vanishing embedding tensor components are
\begin{equation}
\nonumber f_{125} = f_{256} = f_{567} = - f_{157} = -2\,,
\end{equation}
\begin{equation}\label{eq:OurEmbTensor}
\xi_{34}= -3\sqrt 2,\qquad\qquad \xi_{12}=\xi_{17}=-\xi_{26}= \xi_{67} = -\sqrt 2\, k\,,
\end{equation}
and their permutations. For the non-vanishing components of the gravitino shift matrix $P_{\ul{ab}}= -P_{\ul{ba}}$ defined in \eqref{defPab} we find
\bea
P_{12} \!\!&=&\!\! \frac 16 \, e^{-\frac{16}{3}U-\frac 43 V}\left[3\, {\rm Im}(b\,  \overline {c } )+k \right] \,,\qquad P_{34} \,=\,  \frac 16 \, e^{-\frac{10}{3}U-\frac 43 V}\left( 3\, e^{2U} + 2\, e^{2V} \right)  \nn\\ [2mm]
P_{13} \!\!&=&\!\! \frac{1}{2} \, e^{-\frac{10}{3}U-\frac 43 V + \frac 12 \phi}\, {\rm Im}(c  - C_0 b )\,,\qquad 
P_{14} \,=\, -\frac{1}{2} \, e^{-\frac{10}{3}U-\frac 43 V + \frac 12 \phi}\, {\rm Re}(c  - C_0 b )\,, \nn\\ [2mm]
P_{23} \!\!&=&\!\! \frac{1}{2} \, e^{-\frac{10}{3}U-\frac 43 V - \frac 12 \phi}\, {\rm Im}\,b \,, \qquad\qquad\quad\;
P_{24} \,=\, -\frac{1}{2} \, e^{-\frac{10}{3}U-\frac 43 V - \frac 12 \phi}\, {\rm Re}\,b \,.
\eea
This yields
\be
\vct_{\underline a} \,=\, \delta_{\underline a}^5\,,
\ee
so we can choose the su(2)$_\pm$ generators as in footnote~\ref{footnoteSO5gauge}. 

We can then evaluate the constraints for a supersymmetric domain wall found in the general analysis. 
The first condition in \eqref{dsigmavanishes} is trivial, and it turns out that also constraints \eqref{condition_xi}, \eqref{condition_f} are automatically satisfied.
The first in~\eqref{covderAiszero} instead is non-trivial (because of the piece containing the SO(5) connection) and tells that
\be\label{bJcJconstant}
b^J = {\rm const}\,, \qquad c^J = {\rm const}\,.
\ee
The second in~\eqref{dsigmavanishes}, namely $\partial_\sigma Q^r_\pm = 0$, gives
\begin{equation}\label{eq:cbCondition}
c \, =\, (C_0 \pm i\,e^{-\phi})\, b  \,,
\end{equation}
where the two sign choices are correlated.
We also compute $W_\pm$ and find the associated flow equations. One can check that these are compatible with \eqref{bJcJconstant}, \eqref{eq:cbCondition}, and that the equations for the remaining scalars $\{U, V, \phi, C_0, a, b\}$ consistently follow from the reduced superpotential and the reduced metric that are found by plugging~\eqref{bJcJconstant}, \eqref{eq:cbCondition} in the original $W_\pm$ and in the scalar kinetic terms. It is therefore enough to present this reduced superpotential, which reads
\begin{equation}\label{superpotIIBonSE}
W_\pm \,=\, e^{- \frac{4}{3} U- \frac{4}{3} V} + \frac23\, e^{- \frac{10}{3} U + \frac{2}{3} V} + e^{- \frac{16}{3} U- \frac{4}{3} V}\Big( e^{-\phi}\,|b|^2 \mp \frac13 k \Big) \,,
\end{equation}
together with the reduced metric, which is
\bea
g_{xy}d\phi^xd\phi^y|_{\rm red} &=& 8(dU)^2 + \Big(\frac 12 + \,e^{-4U-\phi}|b|^2 \Big) \Big[ e^{2\phi}(dC_0)^2 + (d\phi)^2 \Big]\nn \\ [2mm]
&&+\, 2\, e^{-4U} \Big[ e^{-\phi}|d b|^2 \pm \, {\rm Im}\left(b \,d\overline b\right) dC_0 - \frac 12 e^{-\phi}\, d( |b|^2)\,d\phi \Big]\nn \\ [2mm] 
&&+\, 2\,e^{-8U}\Big[ da+ \frac 12 |b|^2 \, dC_0 \pm e^{-\phi}\,{\rm Im}\left( b \,d \overline b\right) \Big]^2\,.
\eea
Choosing the upper sign in \eqref{sigmaflow}, \eqref{FloweqFromGaugino_MainText}, the associated flow equations read:
\begin{eqnarray}
U' &=& e^{- \frac{10}{3} U + \frac{2}{3} V} + \frac 12\, e^{-\frac{16}{3}U -\frac{4}{3}V}\Big( 3\, e^{-\phi}|b|^2 \mp k  \Big) \,, \nonumber\\  [2mm]
V' &=& 3\,e^{-\frac{4}{3}U -\frac{4}{3}V} -2\,e^{- \frac{10}{3} U + \frac{2}{3} V} + \frac 12\, e^{-\frac{16}{3}U -\frac{4}{3}V}\Big( 3\, e^{-\phi}|b|^2 \mp k \Big) \,, \nonumber\\  [2mm]
b' &=& -3\,e^{-\frac{4}{3}U-\frac{4}{3}V}\, b\,,\nonumber\\ [2mm]
\phi' &=& C_0' \;=\; a' \;=\; 0\,.
\end{eqnarray}
This system can be solved by introducing the new radial coordinate 
\be
d\rho \,=\, e^{-\frac43(U+V)}\,dr\,,
\ee 
so that the equations for $b$ and $(U-V)$ decouple.
 The general solution is
\begin{eqnarray}\label{SEsol}
b&=& b_0\,e^{-3\rho}, \nonumber\\  [2mm]
e^{2(U -V)} &=& 1+\,\alpha \,e^{-6\rho}\,,  \nonumber\\  [2mm]
e^{2(U +V)} &=& \frac{\beta\, e^{4\rho}}{(1 + \alpha\, e^{-6\rho})^{\frac{1}{3}}}   -\frac{|b_0|^2}{e^{\phi}\alpha} + \Big(\frac{|b_0|^2}{e^{\phi}\alpha} \pm \frac{k}{2}\Big)\,{} _{2}F_1\left[1, 1, \tfrac53; -\alpha\,e^{-6 \rho} \right] ,
\end{eqnarray}
where $b_0$ (complex) and $\alpha,\,\beta$ (real) are integration constants. The hypergeometric function appearing in the last line is
\be
_2F_1\left[1,1,\tfrac 53 ;z\right] = \frac{Z}{3\, z} \left[ 2\sqrt 3\left(\frac{\pi}{2} + \,{\rm arctan} \frac{Z - 2}{\sqrt 3\, Z}\right) + \log\frac{1-Z + Z^2}{( 1+  Z)^2}\right] \,,\quad Z = \left( \frac{z}{1-z} \right)^{\frac{1}{3}}\!.
\ee
In the variable $\rho$, the equation for the domain wall warp factor $A$ becomes
\begin{equation}
A^\prime\,=\,1+\frac23\,U' \qquad\Rightarrow\qquad A\,=\,\rho+\frac23\,U\,,
\end{equation}
where in its solution we are neglecting an integration constant, as it can always be reabsorbed in a rescaling of the four-dimensional Minkowski coordinates.

When $\beta=0$, as $\rho\to +\infty$ the scalars take a finite value which extremizes the five-dimensional scalar potential, and the metric is asymptotically AdS$_5$ (with radius $L^4 =|k|/2$). In this case, noting that 
\begin{equation}
\lim_{z\to 0}\,\,_2F_1\left[1, 1, \tfrac 53; z \right]=1\,,
\end{equation}
we see that positivity of $e^{2(U+V)}$ as $\rho\to +\infty$ imposes $k>0$ for the upper sign choice and $k<0$ for the lower choice. Moreover, positivity of $e^{2(U-V)}$ requires $e^{6\rho}> -\alpha$.

We checked that all the five-dimensional equations of motion given in \cite{IIBonSE} are satisfied, as expected from the general analysis in section~\ref{generalDWanalysis}. Due to the consistency of the truncation, the five-dimensional solution lifts to a solution of type IIB supergravity. It is interesting to discuss how it looks like in ten dimensions. From the reduction ansatz given at the beginning of this section, we find that the ten-dimensional metric takes the following D3-brane form
\begin{equation}\label{D3branemetric}
ds^2\,=\,h^{-\frac12}(\rho)\,ds^2(\mathbb R^{1,3})\,+\,h^{\frac12}(\rho)\,ds^2(M_6)
\end{equation}
with warp factor
\begin{equation}\label{10dWarpIIBonSE}
h(\rho)\,=\,\beta + e^{-4\rho}\,\left(1+\alpha\,e^{-6\rho}\right)^{\frac{1}{3}}\,\left[-\frac{|b_0|^2}{e^{\phi}\alpha} + \Big(\frac{|b_0|^2}{e^{\phi}\alpha} \pm \frac{k}{2} \Big) \, {}_{2}F_1\left[1, 1, \tfrac53; -\alpha\,e^{-6 \rho} \right]\right]
\end{equation}
and six-dimensional transverse metric
\begin{equation}\label{CYblownup}
ds^2(M_6)\,=\,\frac{e^{2\rho}}{(1+\alpha\,e^{-6\rho})^{\frac{2}{3}}}\,\Big[ (d\rho)^2\,+\,\eta^2\,+\,(1+\alpha\,e^{-6\rho})\,\,ds^2(\mathcal B_{\rm KE})\Big]\,.
\end{equation}
For $b_0=0$, this solution had already appeared in \cite{PandoZayas:2001iw, BenvenutiEtAl} (see also \cite[app.~B]{UnquenchedKlebanovWitten}).\footnote{The change of variable leading to the expressions given there is $(1+\alpha\, e^{-6\rho})^{-1} = 1- \frac{b^6}{r^6}$, with $b^2=\alpha^{1/3}$.} The metric \eqref{CYblownup}, first found in \cite{PandoZayas:2001iw} in the case of the conifold, is Calabi--Yau and describes a Calabi--Yau cone modified by blowing-up a four-cycle $\mathcal B_{\rm KE}$ \cite{BenvenutiEtAl}. The deformation is controlled by $\alpha$, with the cone being retrieved as $\alpha=0$. In particular, if both $\alpha$ and $\beta$ vanish, the standard AdS$_5\,\times\;$SE$_5$ solution of type IIB supergravity is obtained.
When $b_0\neq 0$, our solution has an extra feature: the complexified three-form of type IIB supergravity is non-trivial and reads
\be
G_3:= i\, e^{\frac\phi2} \left[ie^{-\phi} dB - (dC_2-C_0 dB)\right] \,=\, 3\,b_0\, e^{-\frac{\phi}{2}-3\rho}\,  \Omega\wedge (d\rho - i \eta)\,
\ee
for the lower sign choice in \eqref{10dWarpIIBonSE}, and its complex conjugate for the upper choice. This is closed, imaginary (anti-)self-dual and primitive with respect to the Calabi--Yau structure. So we conclude that the present $\mathcal N=1$ solution falls in the well-known class of supersymmetric warped Calabi--Yau backgrounds of type IIB supergravity with non-trivial three-form flux and constant axio-dilaton, first discussed in \cite{GranaPolchinski1}.
 
We also observe that for $b_0 \neq 0$ one can take a limit which, when the Sasaki--Einstein manifold is $S^5$, corresponds to the SU(3)-invariant sector of the GPPZ flow~\cite{GPPZflow} (see \cite{PilchWarnerLift} for its lift to type IIB supergravity). This is most easily seen by choosing the five-form flux so that $k=2$. Then, with $\beta=0$ and relating the other integration constants as $|b_0|^2 = - \alpha\,e^{\phi}$ (so we need $\alpha <0$), the third line of (\ref{SEsol}) simplifies, and the scalars with a non-trivial profile can be parameterized by the single function
\be
\lambda(\rho) \,= \,\frac{1}{2}\log \left[\frac{1+ (-\alpha)^{1/2} \,e^{-3\rho}}{1- (-\alpha)^{1/2} \,e^{-3\rho}}\right]
\ee
as
\bea
-U \!&=&\! \displaystyle V \;=\; {{\frac{1}{2}}}\log{\left(\cosh{\lambda}\right)}, \nn\\ [2mm]
\left|\frac{c}{C_0 + i\,e^{-\phi}}\right|  \!&=&\! \, |b|\; \;=\; e^{\phi/2}\tanh{\lambda}\,.
\eea
The superpotential reduces to $W= \cosh^2 \lambda$, and the flow equations to $\lambda' = -3\,\partial_\lambda W$. This precisely describes the SU(3)-invariant sector of the GPPZ flow.
We can thus conclude that tuning $|b_0|^2$ from $0$ to $-\alpha\, e^\phi$, our solution interpolates between the solution of~\cite{PandoZayas:2001iw, BenvenutiEtAl} and the one of~\cite{GPPZflow}.

Finally, from the asymptotic behaviour of the scalar fields we can identify which operators in the dual field theory are driving the RG flow. Denoting by $\mathcal W$ the gauge superfield of the $\mathcal N=1$ SCFT dual to type IIB on AdS$_5\times\,$SE$_5$, we find (see for instance \cite{IIBonSE}) that we are switching on a vev for the first components of ${\rm Tr}(\mathcal W^2)$ and of ${\rm Tr}(\mathcal W^2\overline{\mathcal  W}{}^2)$. The former is a relevant operator with conformal dimension $\Delta =3$, which drives the GPPZ flow and is dual to (the fluctuation of) $b -ic$. The latter has $\Delta = 6$, so it is irrelevant, and is dual to $U-V$. Setting $\beta=0$ we have eliminated a possible further deformation, corresponding to an irrelevant source D-term $\int d^2 \theta d^2\bar \theta\:{\rm Tr}({\mathcal W}^2\overline {\mathcal W}{}^2)$, which has $\Delta=8$ and is dual to $4U+V$ (being an irrelevant source in the Lagrangian, this cannot be switched on as long as the CFT is regarded as the UV theory, and correspondingly the asymptotic AdS factor on the gravity side is preserved).

\section{New superpotentials for conifold solutions}\label{dwT11}

In the construction of string theory solutions dual to non-trivial RG flows, a seminal role was played by the conifold \cite{CandelasOssaConifold}. This is a Calabi--Yau three-fold constructed as the cone on the coset space \hbox{$T^{1,1}= [{\rm SU}(2)\!\times\! {\rm SU}(2)]/{\rm U}(1)$}, endowed with its Sasaki--Einstein metric. The superconformal field theory dual to type IIB on AdS$_5\times T^{1,1}$, which describes the low-energy dynamics of D3-branes probing the tip singularity of the conifold, was identified in~\cite{KlebanovWitten}. The addition of fractional D3-branes breaks conformality and induces RG flows with non-trivial properties such as duality cascades, confinement and chiral symmetry breaking \cite{KlebanovNekrasov,KlebanovTseytlin,KlebanovStrassler,PandoZayasTseytlin,BaryonicBranchKS}. Similar features are obtained by wrapping D5-branes on the two-cycle of $T^{1,1}\sim S^2 \times S^3$~\cite{MaldacenaNunez}. In the corresponding type IIB supergravity solutions, the cone geometry gets deformed, in a way that also affects the metric on $T^{1,1}$. All the conifold solutions~\cite{KlebanovTseytlin,KlebanovStrassler,MaldacenaNunez,PandoZayasTseytlin,BaryonicBranchKS} preserve four-dimensional Poincar\'e invariance and can be seen from a five-dimensional perspective as domain walls of the type studied in this paper. Not all the relevant $T^{1,1}$ deformations are captured by the universal Sasaki--Einstein truncation considered in the previous section. However, in \cite{T11reduction} (see also \cite{Bena:2010pr,Liu:2011dw}), a larger $\mathcal N=4$ consistent truncation containing all known conifold solutions was constructed by retaining all (and only) the Kaluza--Klein modes of type IIB supergravity on $T^{1,1}$ which are invariant under the action of ${\rm SU}(2)\times {\rm SU}(2)$. 
This adds an $\mathcal N=4$ vector multiplet to the universal Sasaki--Einstein truncation, and introduces further non-trivial gaugings, which in type IIB correspond to background three-form fluxes threading the three-cycle of $T^{1,1}$.

In the following, we scan for supersymmetric, ${\rm SU}(2)\times {\rm SU}(2)$ invariant conifold solutions by applying the general results of section~\ref{generalDWanalysis} to the $\mathcal N=4$ supergravity model of \cite{T11reduction}. We perform an exhaustive search within a subset of the $\mathcal N=4$ scalars, originally introduced by Papadopoulos and Tseytlin (PT) in~\cite{PapadopoulosTseytlin}, which is sufficient for describing all known conifold solutions. Note that the results of \cite{T11reduction} are anyway essential in order to identify the $\mathcal N=4$ structure of the PT ansatz. Besides recovering  the known superpotentials and flow equations for the solutions in \cite{KlebanovTseytlin,KlebanovStrassler,MaldacenaNunez}, we find two new superpotentials: the one for the solution describing the baryonic branch of the Klebanov--Strassler theory \cite{BaryonicBranchKS}, generalizing the previous ones, and the one for the solution describing supersymmetric D3- and wrapped D5-branes on the resolved conifold.

\subsection{Superpotential for the baryonic branch of Klebanov-Strassler}\label{Wbarbranch}

The PT ansatz~\cite{PapadopoulosTseytlin} (see also \cite{BergHaackMueck} for its formulation in five dimensions) contains nine scalars, denoted by $p$, $x$, $g$, $a$ (from the internal metric), $b$ (from the RR two-form potential), $h_1$, $h_2$, $\chi$ (from the NSNS two-form potential) and $\phi$ (the dilaton). Moreover, it contains two parameters $P$ and $Q$, corresponding to RR three-form and five-form fluxes, respectively.
As in the previous section, we proceed by constructing the $\mathcal N=4$ fermionic shifts using~\cite{T11reduction}, then we evaluate the algebraic constraints, and eventually write down the superpotential and the flow equations, checking compatibility with the constraints. Since in the present case the intermediate steps are considerably more involved than in the previous section, we will just give the results.\footnote{For the ease of comparison with the literature, we give our results adopting the normalizations of \cite{PapadopoulosTseytlin}. It follows that the flow equations read
\vskip -3mm
\begin{equation}
A' \,=\, -\frac{1}{3}W\;,\qquad \qquad \varphi^{a\,\prime}\,=\, \frac{1}{2}G^{ab}\frac{\partial W}{\partial \varphi^b}\,,\nn
\end{equation}
where $\varphi^a$ are the scalars in the PT ansatz and $G_{ab}$ is the kinetic matrix in \cite{PapadopoulosTseytlin}. Once our algebraic constraints are taken into account, the relation between the scalar potential $V$ of \cite{PapadopoulosTseytlin} and the superpotential is 
\begin{equation}
V \,=\, \frac{1}{8}G^{ab}\frac{\partial W}{\partial \varphi^a}\frac{\partial W}{\partial \varphi^b} - \frac{1}{3} W^2\,.\nn
\end{equation}
The scalar in the supergravity multiplet is identified as $\sigma = 2\sqrt 3 \, p$. For the map between the other $\mathcal N=4$ fields and the PT ones, see appendix B of \cite{T11reduction}.
\label{footnotePT}}

For the PT system, constraint~\eqref{condition_xi} is automatically satisfied, while the first condition in~\eqref{dsigmavanishes} demands that either 
\be\label{CondOne}
a^2 + e^{2g} = 1 
\ee
or 
\be\label{CondTwo}
(1+a^2+e^{2g})h_2P + a(2h_1P + Q) = 0\,.
\ee
The first possibility eventually leads to just the warped deformed conifold solution of~\cite{KlebanovStrassler}, so we will not discuss it further. 
Let us instead study condition~\eqref{CondTwo}, which has more interesting consequences. The subcase $a=0$ is special and will be discussed in the next subsection. Here we assume $a\neq 0$ and $P\neq 0$, and use \eqref{CondTwo} to fix $h_1$ in terms of the other fields. Then the second constraint in~\eqref{dsigmavanishes} provides an expression that can be used to determine e.g.~$h_2$.
We find:
\begin{equation}\label{solh1h2}
h_1 = -C \,h_2 - \frac{Q}{2P}\,, \qquad\quad
(h_2)^2 = \frac{ e^{2\phi} P^2 (b\,C -1)^2 - e^{2x-2g+\phi}  (a\,C-1)^2}{S^2}\,,
\end{equation}
where we introduced the following functions of $a$ and $g$
\bea
C &\equiv& \frac{1 + a^2 + e^{2g}}{2a}\,,\label{defC}\\ [2mm]
S &\equiv& \frac{\sqrt{a^4 + 2a^2\left(-1+e^{2g}\right) + \left(1+e^{2g}\right)^2}}{2a}\,,
\eea
which satisfy $C^2-S^2 = 1$. One can see that in this way all constraints in~\eqref{NecSufCond} are satisfied.
Plugging~\eqref{solh1h2} back into $W$ computed as in~\eqref{eq:ExprW}, we obtain the following superpotential:
\begin{equation}\label{Wnew}
W \,=\, e^{-2p-2x-g} a\,S \,+\, e^{4p}\,S^{-1}\left[C  + e^{-2x+\phi}P^2 \left(b-C\right)\left(b\,C - 1 \right) \right].
\end{equation}
One can check that via the relation in footnote~\ref{footnotePT} this reproduces the PT scalar potential with $h_1$ and $h_2$ integrated out as dictated by \eqref{solh1h2}.
The flow equations following from~\eqref{Wnew} read
\bea
C' &=& S \nn\\ [2mm]
b' &=& \frac{1-b\,C}{S}\nn \\ [2mm]
a' &=&  \frac{1- a\,C}{S}\,e^{g-6p-2x} + \frac{a(a-b)S}{b\,C-1}    \nn \\ [2mm]
x' &=& a\,S\, e^{-g-6p-2x} + \frac{b-C}{2(b\,C -1)S}\left[ e^{-2g}(a\,C-1)^2  + 2\,e^{-2x-\phi}(h_2)^2S^2 \right]\nn 
\eea
\bea
6\, p' &=& a\,S\, e^{-6p-2x-g} - \frac{2(b-C)}{(b\,C -1)S}\left[ e^{-2g}(a\,C-1)^2  + e^{-2x-\phi}(h_2)^2S^2 \right] - \frac{2C}{S} \nn\\ [2mm]
\phi'&=&   \frac{(C-b)\left(a\,C-1 \right)^2}{\left(b\,C-1 \right)S}\,e^{-2g}\nn \\ [2mm]
\chi' &=& \frac{2a\left(b-C\right)\left(a\,C-1\right)}{b\,C-1 }\,e^{-2g}h_2 \,S 
\,.\eea
Here, the derivative is with respect to the rescaled radial coordinate $dt = e^{4p}du$, where $u$ is the PT coordinate. Moreover, we traded $g$ for $C$ using \eqref{defC}, and $h_2$ is to be replaced by~\eqref{solh1h2}. The kinetic matrix used in the formula for the flow is the one obtained by plugging~\eqref{solh1h2} in the kinetic terms of \cite{PapadopoulosTseytlin}.
The equations above are precisely those for the solution describing the baryonic branch of the Klebanov--Strassler theory, which were obtained in \cite{BaryonicBranchKS} by other means, working in ten dimensions.\footnote{When comparing with \cite[app.~B]{BaryonicBranchKS}, one has to recall that they use the string frame, while as \cite{PapadopoulosTseytlin} we work in the Einstein frame. Comparing the respective ten-dimensional metric ans\"atze, one finds $x_{\rm there} = x_{\rm here} + \phi/2$, $p_{\rm there} = p_{\rm here} - \phi/6$ and $A_{\rm there} = (A+p-x/2+\phi/4)_{\rm here}$. Moreover, in \cite{BaryonicBranchKS} the three-form flux parameter was fixed to $P=-1/2$, while we leave it arbitrary.} 
Hence we have found a superpotential for that solution, which was unknown before. This  provides a monotonic $c$-function along the flow.

Starting from our expression \eqref{Wnew}, the known superpotentials for the Maldacena--Nu\~nez \cite{MaldacenaNunez} and the Klebanov--Strassler \cite{KlebanovStrassler} solutions can be recovered as limiting cases once one restricts to the respective systems of variables. The Maldacena--Nu\~nez variables are obtained from the PT ones by setting \cite{BergHaackMueck}:
\begin{equation}
\chi = h_1 = h_2 = Q= 0\,,\,\quad b=a\,,\quad\, \phi = -6p-g-2\log P\,, \quad\, x = \frac 12 g - 3p\,,
\end{equation}
so \eqref{solh1h2} are consistently satisfied and $W$ becomes
\begin{equation}
W_{\rm MN} = 2\,e^{4p-2g}a\, S \,=\, e^{4p}\sqrt{\left(a^2-1\right)^2e^{-4g} + 2\left(a^2+1\right)e^{-2g} + 1}\,,
\end{equation}
which is the superpotential of the Maldacena--Nu\~nez solution, found in \cite{PapadopoulosTseytlin}.
In order to recover the Klebanov--Strassler system, we parameterize $a$ and $g$ as 
\begin{equation}\label{PTtoKS}
a = \tanh y\,,\qquad e^{-g} = \cosh y\,,
\end{equation}
so that $C= \coth y$, and \eqref{solh1h2} becomes
\begin{equation}\label{solh1h2KS}
h_1 = -h_2\coth y - \frac{Q}{2P}\,,\qquad\quad
(h_2)^2 = e^{2\phi} P^2 \,(b \cosh y - \sinh y)^2\,,
\end{equation}
while $W$ in \eqref{Wnew} reads
\begin{equation}
W_{\rm KS}|_{h_1,h_2} \,=\, e^{-2p-2x} + e^{4p}\cosh y + e^{4p-2x+\phi} P^2 \left(b \cosh y - \sinh y \right)\left(b- \coth y\right).
\end{equation}
This is the known superpotential for the Klebanov--Strassler solution, namely \cite{PandoZayasTseytlin}
\begin{equation}
W_{\rm KS} = e^{-2p-2x} + e^{4p}\cosh y - \frac 12 e^{4p-2x}(Q+ 2P h_1+ 2P\, b \,h_2)\,,
\end{equation}
with $h_1,\,h_2$ replaced by \eqref{solh1h2KS}.

Another limiting case that can be considered is the one in which both the NSNS three-form and the RR five-form vanish, yielding solutions describing just wrapped D5-branes. This setup was studied in \cite{CaseroNunezParedes}, and a superpotential for it was found in \cite[sect.~3]{HoyosNunezPapadim}. In the PT ansatz it corresponds to the truncation $h_2 = 0$, $h_1 = -{Q}/{(2P)}$ and $\chi = {\rm const}$. Using the second in \eqref{solh1h2}, one can check that then our superpotential reduces to the one in~\cite{HoyosNunezPapadim}.\footnote{The change of variables between \cite{HoyosNunezPapadim} and PT is
\bea
&&P_{\rm HNP}\,=\,4\,a\,S\,e^{x-g-\frac{\phi}{2}}\,,\qquad\qquad\qquad Q_{\rm HNP}\,=\,4\,\left(a\,C-1\right)\,e^{x-g-\frac{\phi}{2}}\,,\nonumber\\
Y \!\!\!&=&\! e^{-6p-x-\frac{\phi}{2}}\,,\qquad\qquad \sinh{\tau}\,=\,S^{-1}\,,\qquad\qquad N_c\,=\,4P\,,\qquad\qquad N_f = 0\,,\nn
\eea
and their constraint $\omega = 0$ is the same as the second in our~\eqref{solh1h2}, with $h_2=0$.
}

\subsection{Supersymmetric D3  and D5-branes on resolved conifold}

In the following we study the case $a=0$, which was left aside in the previous section and allows for a different solution of the constraints leading to a different superpotential. Geometrically, $a=0$ means that there are no mixed terms in the metric on the $\mathbb{CP}^1 \times \mathbb{CP}^1$ base of the U(1)-fibration describing $T^{1,1}$. With $a=0$, constraint \eqref{CondTwo} implies $h_2 =0$ (we are interested in solutions with $P\neq 0$). Then the remaining algebraic conditions of section~\ref{NecSufCond} also require $b=0$. We are thus led to consider an $a=b=h_2=0$ truncation of the PT ansatz, which corresponds to the ansatz first considered by Pando-Zayas and Tseytlin in \cite{PandoZayasTseytlin} in order to study solutions based on the resolved conifold geometry.\footnote{The change of variables leading to the notation in \cite{PandoZayasTseytlin} is $h_1=\frac{1}{2}(f_1-f_2)$, $\chi = \frac{1}{2}(f_1+f_2)$, $g=y$.}
The type IIB supergravity solution found there, describing regular D3 and wrapped D5-branes on the resolved conifold, is known to be non-supersymmetric \cite{Cvetic:2000db} (see also \cite{GwynKnauf}), although it can be described in terms of a simple superpotential. In the following, applying our general method we identify a supersymmetric solution describing D3 and wrapped D5-branes on the resolved conifold, having the same active fields as in~\cite{PandoZayasTseytlin}, plus the dilaton.
For the superpotential we find
\begin{equation}\label{PZ-TSup}
W= e^{-2p-2x}\cosh g + e^{4p} + \frac 12 e^{4p-2x} \sqrt{4\,e^{2x+\phi}P^2\sinh^2g + \left(Q + 2P h_1\right)^2}\,.
\end{equation}
This is similar to the one given in \cite{PandoZayasTseytlin}, the latter being recovered by erasing the first term in the square root. Both superpotentials define BPS domain walls in that both reproduce the same truncated PT potential $V|_{a=b=h_2=0}$ through the relation in footnote~\ref{footnotePT}. However, only the one in \eqref{PZ-TSup} generates a supersymmetric flow. Note in particular that the new term introduces a dependence on the dilaton; as a consequence, $\phi$ will flow in our solution.

We present the system of first-order equations for the fields $\{x,\,g,\,p,\,\phi,\,h_1,\,\chi\}$ following from (\ref{PZ-TSup}) in the radial coordinate $\rho$, related to the PT one by $d\rho = e^{4p}du$. We find:
\begin{eqnarray}\label{FloweqsResConrho}
x'&=&  -e^{-6p-2x}\cosh g - \frac{1}{2} e^{-2x}\,T^{-1} \left[ (Q + 2Ph_1)^2 + 2\,e^{2x+\phi}P^2\sinh^2 g \right] \nonumber \\ [2mm]
g'&=& e^{-6p-2x}\sinh g + e^{ \phi} P^2 T^{-1} \sinh(2g) \nonumber\\ [2mm]
\phi' &=& 2\,e^{\phi}P^2 T^{-1}\sinh^2 g \nonumber\\ [2mm]
p'&=&  \frac{1}{3} - \frac{1}{6}\, e^{-6p-2x}\cosh g +  \frac{1}{6} \, e^{-2x} \,T  \nonumber\\ [2mm]
h_1'&=& e^{ \phi}P\, T^{-1}\cosh(2g) \,(Q+2 Ph_1)  \nonumber\\ [2mm]
\chi'&=&e^{ \phi}P\, T^{-1}\sinh(2g) \,(Q+2 Ph_1)  \,, 
\end{eqnarray}
where we introduced the quantity
\begin{equation}\label{defT}
T \,=\, \sqrt{(Q +2 P h_1)^2 + 4\,e^{2x+\phi}P^2\sinh^2g} \,.
\end{equation}
Setting the three-form flux $P=0$, one recovers the equations describing regular D3-branes on the resolved conifold \cite{PandoZayasTseytlin}. 
 Linearizing for small~$P$, which physically means taking a small ratio between the three-form and five-form Page charges, we recover precisely equations (5.15)--(5.18) of~\cite{PandoZayasTseytlin}. However, at higher orders the solution deviates from the one in~\cite{PandoZayasTseytlin}.

The system can be partially solved
by first noting that
\begin{equation}\label{solForT}
(e^{-\phi}\,T)' \,= \, 2P^2\qquad\qquad \Rightarrow \qquad\qquad e^{-\phi}\,T \,=\, 2P^2\rho\,,
\end{equation}
where we are absorbing the integration constant in a shift of $\rho$. This eliminates $T$ from the equations. 
We also introduce the new convenient variables $w=x - \phi/2$ and $z= x+3p$.
The solution for $\phi$ is 
\begin{equation}\label{SolPhi}
\phi  \,=\, \frac{z}{2} + \frac 34 \log \left|\frac{\sinh g}{2P^2\rho}\right| -\frac \rho2 + \gamma\,,
\end{equation}
with $\gamma$ an integration constant. Moreover we find that $w$ is determined by
\be\label{Solw}
e^{2w}\sinh^2 g\,=\,\left(1 -\kappa^2 \, e^{2\phi}\right) P^2\rho^2\,,
\ee
where $\kappa$ is a constant such that $\kappa^2 \leq e^{-2\phi}$, and that $h_1$ is fixed by
\begin{equation}\label{solforF5}
 Q + 2P h_1 \,=\, 2\kappa \,P^2\,e^{2\phi} \rho \,.
\end{equation}
The warp factor reads
\begin{equation}\label{SolWarp}
A \,=\,  \frac{1}{3}(\rho - w) \,.
\end{equation}
We are thus left with three equations:
\begin{eqnarray}\label{FinalSys}
g'&=& e^{-2z}\sinh g + \frac{\sinh(2g)}{2\rho}\,, \nonumber\\ [2mm]
z'&=&  1 - \frac{3}{2}\, e^{-2z}\cosh g  + \frac{\sinh^2 g}{2\rho}\,,\nonumber\\[2mm]
\chi'&=& \kappa \, P\, e^{2\phi} \sinh(2g) \,,
\end{eqnarray}
where just the first two need to be solved since knowing the derivative of $\chi$ in the PT ansatz is enough to determine the ten-dimensional background. 

This reduced system is not new. From \eqref{solforF5} we observe that $\kappa = 0$ implies $Q+ 2Ph_1=0$, which in the present ansatz means that the type IIB five-form $F_5$ vanishes, i.e.\ there is no D3-brane charge and we are left just with a D5-brane setup. In this case, the equations have already been studied in \cite{CaseroNunezParedes}. The solution for $\kappa \neq 0$ is instead a particular case of the one discussed in \cite{MaldacenaMartelli} when one takes $a=b=0$. In \cite{MaldacenaMartelli}, a U-duality transformation was proposed mapping a solution with just D5 charge into a solution with both D3 and D5-charge. If we denote by $w_0$ the solution (\ref{Solw}) in the absence of D3-brane charge and thus with $\kappa=0$, then the general solution reads
\be
e^{2w}\,=\,\left(1-\kappa^2\,e^{2\phi}\right)\,e^{2w_0}\,.
\ee
This is precisely the form of the transformation described in \cite{MaldacenaMartelli, Elander}, where the dilaton and the metric functions $g$, $z$ are unaltered. Moreover, our D3-brane charge \eqref{solforF5} and the $\chi$ equation also agree with their formulae. We conclude that any solution to the system (\ref{FloweqsResConrho}) can be obtained from a solution with just D5-brane charge via the transformation of~\cite{MaldacenaMartelli}.

Note that the equations can easily be solved analytically if the special solution \hbox{$g=0$} is chosen, meaning that the two $\mathbb{CP}^1$'s in $T^{1,1}$ have the same size. This leads to the superpotential and the equations given by Klebanov and Tseytlin in \cite{KlebanovTseytlin}. In this case $\chi$ and the dilaton $\phi$ are constant and the general solution for $z$ is
$e^{2z}\,=\,\frac32\,\left(1+z_0\,e^{2\rho}\right).$  
Selecting the integration constant as $z_0=0$ one recovers the solution of~\cite{KlebanovTseytlin} describing regular and fractional D3-branes on the singular conifold, whereas for $z_0\ne0$ the solution was given in~\cite{PandoZayas:2001iw}. Here we explicitly exhibit its supersymmetric nature. The transverse geometry is the same Calabi--Yau as in (\ref{CYblownup}), while the warp factor $h$ is more complicated than the one in section~\ref{dwIIBonSE}, though still expressed in terms of hypergeometric functions~\cite{PandoZayas:2001iw}. 

It is interesting to look at the form taken by the constraints on the spinorial susy parameter discussed in section \ref{NecSufCond}. These can be lifted to IIB supergravity using the map between ten-dimensional and five-dimensional fermionic variations derived in \cite{Liu:2011dw}.
In their spinor conventions, we find that the type IIB supersymmetry parameters are constrained as\footnote{In five-dimensional language we have
\begin{eqnarray}
\varepsilon- \gamma^5\,\sigma_3\,\varepsilon&=& 0\,,\nonumber \\[2mm]
\varepsilon - \gamma^5(\cos\beta \, \varepsilon  -  \sin\beta\,\sigma_1\,\varepsilon^c) &=& 0\,,\nonumber
\end{eqnarray}
where $\varepsilon = {\varepsilon_1 \choose \varepsilon_2}$ is an Sp(2) doublet of Dirac spinors as in \cite{Liu:2011dw}. In this notation, the type IIB Weyl supersymmetry parameter is given by $\epsilon_{\rm IIB} = \varepsilon_1\otimes \eta + \varepsilon_2 \otimes \eta^c$, with $\eta$ being a Killing spinor on the $T^{1,1}$ and $\eta^c$ its charge conjugate.}
\begin{eqnarray}
\epsilon-\frac18\,\Gamma^{0123}\,\Gamma^{AB}\,J_{AB}\,\epsilon&=&0\,,\nonumber\\ [2mm]
\epsilon -i\Gamma^{0123}\left(\cos\beta\,\epsilon + \tfrac{1}{8}\,\sin\beta\,\Gamma^{AB}\,J_{AB}\,\epsilon^c \right)  &=&0\,,
\end{eqnarray}
where $\cos\beta\,=\,\kappa\,e^\phi$. The first projector is compatible with D5-branes filling the four flat spacetime directions and wrapping the submanifold in $T^{1,1}$ identified by the K\"ahler form $J$ on $\mathbb{CP}^1 \times \mathbb{CP}^1$. The second projector describes D3-branes filling the four flat spacetime dimensions and polarized into the D5-branes through the Myers effect \cite{MyersDielectric}; it is of the type considered in \cite{PilchWarner2004}.  

\begin{figure}[!htbp]
\centering{
\includegraphics[scale=.90]{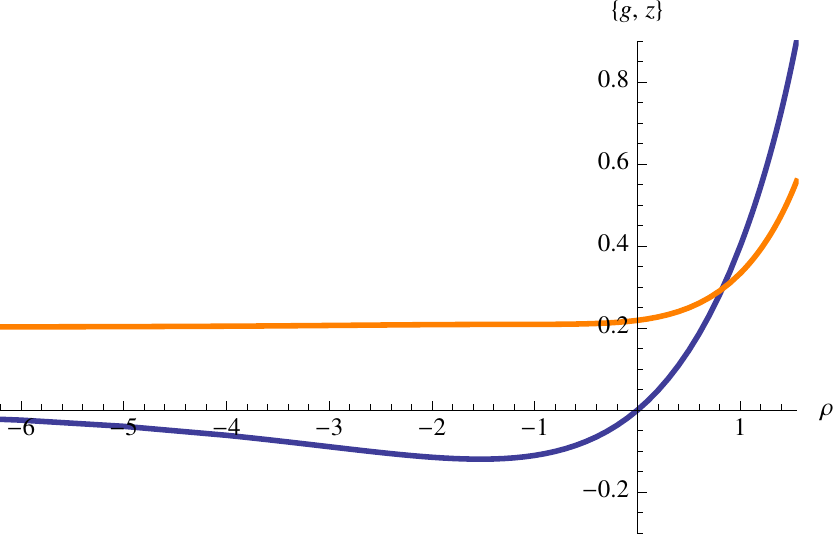}
}
\caption{\label{plots1} Metric functions $g(\rho)$ (dark blue) and $z(\rho)$ (light orange) for the initial values $g(1)=2/5$, $z(1)=1/3$. For the fluxes, we chose $P=Q=1$. The functions asymptote to $g=0$ and $z=\frac12\log{\frac32}$ when $\rho\to-\infty$, which are the values of the singular conifold solution in \cite{KlebanovTseytlin}. The singularity appears at $\rho\approx1.55$.}
\end{figure}

We conclude this section by presenting in figure~\ref{plots1} a numeric solution to the $g$ and $z$ equations in (\ref{FinalSys}). We start the integration at an arbitrary point $\rho_0$, using a power series solution around it as initial condition. The system is then integrated both to the left and right of the starting point. 
We have not explored thoroughly the space of parameters, but an IR singularity appears in all the solutions we generated. This is consistent with the vanishing of $a$ and $b$ modes that are crucial for avoiding the singularity. In the UV, corresponding to $\rho\to-\infty$, the behaviour of the conifold solution in \cite{KlebanovTseytlin} is recovered, and at first-order the metric functions $g$ and $z$ go as
\be
 g\,\sim\,\rho^P\,e^{\frac23\rho}\,,\qquad\qquad\qquad\qquad z\,\sim\, \tfrac12\log{\tfrac32} + e^{2\rho}\,.
\ee 
The transverse space is not Ricci-flat when evaluated in our numeric solutions.
Furthermore, the scalar curvature contains an IR singularity.

\section{Conclusions}\label{Conclusions}

Domain walls in supergravity theories are holographically dual to field theory RG flows and are thus interesting solutions to explore.
In five-dimensional, $\mathcal{N}=2$ supergravity the problem of constructing BPS domain walls was addressed some time ago \cite{CveticDW,5dN=2domainwalls} and used for instance to recover the superpotentials of some conifold solutions \cite{HalmagyiLiuSzepietowski}. However, the $\mathcal{N}=2$ approach is in this case limited and one needs to go to $\mathcal{N}=4$ in order to supersymmetrize the complete PT ansatz and all the solutions it encompasses.

In this paper, we have presented a compact set of necessary and sufficient conditions for supersymmetric domain wall solutions to $\mathcal{N}=4$ gauged supergravity in five dimensions with vanishing $\xi_M$ components of the embedding tensor. As summarized in section~\ref{NecSufCond}, these take the form of first-order flow equations for the warp factor and the scalar fields, generated by a superpotential and, crucially, completed by a set of algebraic constraints involving the scalars and the embedding tensor which specifies the gauging of the model. We found that supersymmetric domain walls are generically 1/4 BPS, namely they preserve four supercharges. The peculiar 1/2 BPS case in which the vector $\widetilde X$ defined in (\ref{Xdefinition}) vanishes was only partially considered; we postpone its complete study to future work.

We subsequently applied our general conditions to two particular $\mathcal N=4$ consistent truncations of type IIB supergravity (so the solutions are automatically embedded into string theory). The first is based on a reduction on an internal manifold with Sasaki--Einstein structure \cite{IIBonSE, GauntlettVarelaIIBonSE}. We gave the explicit solution for its most general supersymmetric domain wall, thus describing a universal RG flow deforming superconformal theories on \hbox{D3-branes} at Calabi--Yau cones. Its ten-dimensional uplift contains a primitive imaginary-self-dual three-form; tuning the associated parameter we interpolate between the solution in \cite{PandoZayas:2001iw, BenvenutiEtAl} and the one in \cite{GPPZflow}.
The second model is an extension of the first for the particular case of $T^{1,1}$, and furnishes an $\mathcal{N}=4$ supersymmetrization of the PT ansatz \cite{T11reduction, Bena:2010pr}. As such, all known conifold solutions can be obtained within it.
In this framework, we found a superpotential for the solution describing the baryonic branch of the Klebanov--Strassler theory, which encompasses the previously known superpotentials and reduces to them when appropriate limits are taken. We also obtained a superpotential for the solution corresponding to D3 and wrapped D5-branes on the resolved conifold. 

For our construction to work and give supersymmetric solutions, it is essential that the constraints in section~\ref{NecSufCond} are satisfied. In the concrete examples we studied, the constraints turned out to be compatible with the flow equations. They were solved for some of the scalar fields, which were subsequently integrated out, so that a superpotential and a system of flow equations for a smaller set of fields was obtained. It would be useful to prove that this can be done in more generality.

Regarding technical extensions to our work, it would be interesting to repeat the general analysis by also including the $\xi_M$ components of the $\mathcal N=4$ embedding tensor \cite{SchonWeidner}. Another obvious development would be to study charged domain walls by allowing a non-trivial profile for some of the $\mathcal N=4$ supergravity form fields. We have restricted our attention to the Poincar\'e invariant case, but domain walls with curved four-dimensional leaves (mainly AdS) are also of interest (see \cite{CurvedDWI, CurvedDWII} for the $\mathcal{N}=2$ case). Finally, the strategy developed in this paper should be smoothly translatable to the case of four-dimensional, $\mathcal{N}=4$ gauged supergravity and applicable e.g.\ to the truncations in \cite{TriSasaki}.

In addition to those discussed in sections~\ref{dwIIBonSE} and \ref{dwT11}, there exist other five-dimensional $\mathcal N=4$ gauged supergravity models that might contain supersymmetric domain wall solutions with a string theory lift. For instance, a simple possibility would be to consider pure supergravity with its inequivalent ${\rm SU}(2)\times {\rm U}(1)$ gaugings \cite{RomansMagnetovac}, and study the possible dilatonic domain walls with a non-trivial profile for the scalar in the gravity multiplet, lifted according to \cite{Lu:1999bw, Gauntlett:2007sm}. Other appealing examples are the $\mathcal{N}=4$ supergravities considered in \cite{OrbifoldN=4} and related to reductions on orbifolds of the sphere. 

Finally, it would be interesting to explore the possibility of deforming the superpotential for supersymmetric domain walls to a fake superpotential for non-supersymmetric solutions. It should also be possible to employ the superpotential presented in section \ref{Wbarbranch} to study non-supersymmetric deformations to the solution in \cite{BaryonicBranchKS} by using the computational technique presented in \cite{Borokhov}.

\section*{Acknowledgments}

\noindent We would like to thank Dario Martelli and Carlos N\'u\~nez for interesting discussions. The work of D.C. is supported by an STFC grant ST/J002798/1. The work of G.D. is supported in part by the ERC Advanced Grants no. 226455, ÒSUPERFIELDSÓ, by the European Programme UNILHC (contract PITN-GA-2009-237920), by the Padova University Project CPDA105015/10, by the MIUR-PRIN contract 2009-KHZKRX and by the MIUR-FIRB grant RBFR10QS5J. The work of A.F. is supported by STFC grant ST/J00040X/1.

\appendix

\section{More on the scalar manifold geometry}\label{appScalarManifoldGeom}

The SO(5) gamma matrices $\Gamma_{\ul{a}}{}_i{}^j$ satisfy
\begin{equation}
\Gamma_{\underline a}{}^{ij}= -\Gamma_{\underline a}{}^{ji}\;,\qquad \Gamma_{\underline a}{}_{\,i}{}^{i} = 0\;,\qquad \left(\Gamma_{\underline a}{}^{ij}\right)^* = \Omega_{ik}\Omega_{jl}\Gamma_{\underline a}{}^{kl}\,,
\end{equation}
as well as
\be
\{\Gamma^{\ul{a}},\Gamma^{\ul{b}}\} \,=\, 2 \,\delta^{\ul{ab}} \bbone_4\,,
\ee
\begin{equation}
\Gamma^{\underline a\, ij} \Gamma_{\underline a\, kl} \,=\, 4\,\delta^{ij}_{kl}- \Omega^{ij} \Omega_{kl}\,,
\end{equation}
and the relations between antisymmetric products
\begin{equation}\label{Gamma=*Gamma}
\Gamma_{\underline{a_1\ldots a_k}} \;=\; \frac{(-1)^{[k/2]}}{(5-k)!}\epsilon_{\underline{a_1\ldots a_5}}\Gamma_{\underline{a_{k+1}\ldots a_5}}\qquad\qquad \textrm{(with $\epsilon_{12345}=+1$)}\;.\\[2mm]
\end{equation}
It is also useful to record that
\be
\Gamma_{\ul{ab}}{}_{\,ij} \Gamma^{\ul{cd}}{}^{\,ij} \,=\, 8\,\delta^{\ul{cd}}_{\ul{ab}}\,,
\ee
so for any antisymmetric tensor $T_{\ul{ab}}$ the relation between its components in vectorial and in spinorial indices is
\be
T_{ij} = T_{\ul{cd}}\Gamma^{\ul{cd}}{}_{\,ij} \qquad \Leftrightarrow \qquad  T_{\ul{ab}} = \frac 18\Gamma_{\ul{ab}}{}_{\,ij}\, T^{ij}\,. 
\ee

The $\frac{{\rm SO}(5,n)}{{\rm SO}(5)\times {\rm SO}(n)}$ coset representative $(\mathcal V_M{}^{\underline a},\mathcal V_M{}^{a})$ and its inverse $(\mathcal V_{\underline a}{}^M,\mathcal V_{a}{}^M)^T$ satisfy
\begin{eqnarray}
	{\cal V}_{\underline a}{}^{M} {\cal V}_M{}^a  &=& 0 \;=\; {\cal V}_{a}{}^{M} {\cal V}_M{}^{\underline a}, \nn\\
	{\cal V}_a{}^{M} {\cal V}_M{}^b  &=& \delta_a{}^b,\nn\\
	{\cal V}_{\ul{a}}{}^M {\cal V}_M{}^{\ul{b}} &=& \delta_{\ul{a}}{}^{\ul{b}} \,.
\end{eqnarray}
It follows that from
\be
\eta_{MN} \,=\, -{\cal V}_M^{\underline a} {\cal V}_N^{\underline a}  + {\cal V}_M^a {\cal V}_N^a \,=\, -{\cal V}_M^{ij} {\cal V}_{N \, ij}  + {\cal V}_M^a {\cal V}_N^a 
\ee
we also have
\begin{equation}
	{\cal V}_M{}^a =  \eta_{MN} {\cal V}^{Na}, \qquad 	{\cal V}_M{}^{\underline a} =- \eta_{MN} {\cal V}^{N\underline a}\,.
\end{equation}

For the covariant derivative $D_x$ with respect to the SO$(5)\times$SO$(n)$ composite connection on the scalar manifold, one has
\begin{eqnarray}	
	D_x {\cal V}_{M}{}^{ij} &=& -\frac12 {\cal V}_M{}^a \vielb_{x}^{a\,ij}\, \,,\qquad\qquad
	D_x {\cal V}^M{}_{ij} \,=\, +\frac12 {\cal V}^{Ma} \vielb_{x\,ij}^a\,, \label{eq:DxV_Mij}\\ [2mm]
	D_x {\cal V}_{M}{}^a &=& -\frac12 {\cal V}_M{}^{ij} \vielb_{x\,ij}^a\,, \qquad\qquad
	D_x {\cal V}^{M a} \,=\, +\frac12 {\cal V}^{M}{}_{ij} \vielb_{x}^{a\,ij}\,.
\end{eqnarray}

Finally, the curvature two-form of the SO(5) connection, 
\be\label{RfromConnection}
R^{\ul{ab}}= \frac 14(d\omega^{\ul{ab}} + \omega^{\ul{a}}{}_{\ul{c}}\wedge \omega^{\ul{cb}})\,,
\ee
can be computed from \eqref{connfromMCform} using \eqref{eq:DxV_Mij}, \eqref{vielbfromMCform}, and reads
\be
R^{\ul{ab}} = \frac{1}{8} \vielb^{a\ul a} \wedge \vielb^{a\ul{b}}\,.
\ee
Introducing $R_{ij} = R^{\ul{ab}}(\Gamma_{\underline{ab}})_{ij} $, this can be written as
\begin{equation}
	R_{ij} \,=\, 
	\frac18 \, \vielb^a{}_{\,i}{}^k\wedge \vielb^a{}_{kj}.
\end{equation}
or, in terms of its components $R_{ij} = \frac{1}{2} R_{xy\,ij} d\phi^x \wedge d\phi^y$, as
\begin{equation}\label{CurvatureComponents}
	R_{xy\,ij} = \frac14 \, \vielb^a_{[x\,i}{}^k \vielb_{y]kj}^a = \frac{1}{4}\, \vielb^{a \underline a}_x \vielb^{a\underline b}_y\,  (\Gamma_{\underline{ab}})_{ij}\,.
\end{equation}
So we have for the following product of two vielbeine:
\begin{equation}\label{productvielb}
	\vielb^a_{x\,i}{}^k \vielb^a_{y\,k}{}^j \,=\, g_{xy} \delta_i{}^j + 4 R_{xy\,i}{}^j\,.
\end{equation}
Moreover, one can check that the following identity holds
\be\label{RRcomplete}
R_{xy\, ij} R^y{}_{z}{}^{kl} \,=\,  \frac{1}{8} \vielb^{a}_{x\,(i}{}^{(k} \vielb^{a}_{z\,j)}{}^{l) }  - \frac{1}{8}g_{xz}\delta_i^{(k}\delta^{l)}_j  - \frac{1}{2}R_{xz\,(i}{}^{(k} \delta{}^{l)}_{j)} \,.
\ee
By contracting with $\delta^{j}_k$, this becomes
\begin{equation}\label{RR=g+R}
	4 R_{xy \,i}{}^jR^y{}_{z\,j}{}^{k} \,=\, g_{xz} \delta_i{}^k + 3 R_{xz\,i}{}^k.
\end{equation}

\section{Analysis of the gaugino equation}\label{AnalysisGaugino}

In this appendix, we analyse the gaugino equation~\eqref{GauginoEq}, namely
\be\label{GauginoEqAppendix}
 i\, \phi^{x\prime}  \vielb^a_{x\,ij} \gamma_5\epsilon^j  \,+\, 2g\, P^a{}_i{}^j\epsilon_j\;=\; 0\,,
\ee
where the gaugino shift matrix $P^a{}_{\!ij}$ was defined in~\eqref{gauginoshift}.
We prove that when $W_+$ is chosen, and constraint \eqref{projectionspinors} on the supersymmetry parameter is taken into account, this gaugino equation is equivalent to the scalar flow equation
\begin{equation}\label{FloweqFromGaugino}
\phi^{x\,\prime}\,=\,\mp\,3g\,g^{xy}\,\partial_y W_+\,
\end{equation}
(where the sign choice is the same as the one in \eqref{projectionspinors}), together with the algebraic conditions
\bea\label{condition_xi_appendix}
&&\xi^{a\ul{a}} \vct_{\ul a} = 0\,,\\ [2mm]
&&f^{a\ul{ab}} L^r_{+ \ul{ab}}\left(\delta^{rs} + Q_+^r Q_+^s \right) \,=\,0\,,\label{condition_f_appendix}
\eea
which are just eqs.~\eqref{FloweqFromGaugino_MainText}, \eqref{condition_xi} and \eqref{condition_f} in the main text. 
The same statement holds when $W_+$ and $Q_+$ are replaced with $W_-$ and $Q_-$ (we omit the proof since it is completely analogous to the $W_+$ case).

\subsection{Rewriting the equation}

As a first thing, we rewrite the gaugino equation in a more convenient way.
We introduce the following vectors on the vector multiplet scalar manifold:
\be
K^x \; :=\;-\frac{1}{2}\,\Sigma^2\,\vielb^{x\,a\,ij}\,\xi^{MN}\,\mathcal{V}_{M}{}^a\,\mathcal{V}_{N\,ij}
\ee
and
\be
K^x_M \;:=\; -\,\frac{1}{2}\,\Sigma^{-1}\,\vielb^{x\,a\,ij}\,f_M{}^{NP}\,\mathcal{V}_{N}{}^a\,\mathcal{V}_{P\,ij} 
\,. \label{fDef}
\ee
One can show that these vectors are Killing \cite{DallHerrZag}, and that the isometries they generate are precisely those being gauged in the supergravity theory specified by the embedding tensor components $\xi^{MN}$, $f^{MNP}$. 
With these definitions, it is easy to see that the gaugino shift matrix $P^a_{\!ij}$ given in~\eqref{gauginoshift} can be written as\be\label{gauginoshiftwithonlyK}
P^{a}{}_{\!ij} \;=\; \vielb^{a}_{x\,i}{}^k\left(-\frac{1}{2\sqrt{2}}\,\Omega_{kj}K^x-\frac{1}{2}\,K^{Mx}\mathcal{V}_{M\,kj}\right).
\ee
Then the gaugino equation~\eqref{GauginoEqAppendix} is equivalent to
\begin{equation}\label{gauginoEqBIS}
\vielb^a_{x\,i}{}^k\left(i\,\phi^{x\,\prime}\,\delta_k{}^j\gamma_5 +  \frac{g}{\sqrt 2}\delta_k{}^j K^x +g\,  K^{Mx} \mathcal{V}_M{}_k{}^{j}\right)\epsilon_j\,=\,0\,,
\end{equation}
which is the expression we are going to analyse below.

We also note that, starting from the definitions of the various objects and recalling~\eqref{eq:DxV_Mij}, one can prove the following general identity involving the USp(4) covariant derivative of the gravitino shift matrix, the USp(4) curvature and the Killing vectors:
\be\label{DPisRK}
3\,D_xP_{ij}\,=\,-2\sqrt 2\,R_{xy\,ij}\,K^y\, - 4\,R_{xy\,(i|}{}^k\,\mathcal{V}_{M\,k|j)}\,K^{My}\,.
\ee
It also holds that
\be\label{identitycommutator}
K^{My} R_{xy}{}_{[i|}{}^k \mathcal V_{M\, k|j]}\,=\, \frac{1}{4} K_x^M \mathcal V_{M\,ij} \,,
\ee
which can be used to rewrite the last term in~\eqref{DPisRK} in various ways.\footnote{Though we will not need this in our proof, let us note that using identities~\eqref{DPisRK}, \eqref{identitycommutator}, the gaugino shift matrix can be recast in a more general form. Multiplying \eqref{DPisRK} with a vielbein and using~\eqref{identitycommutator} gives
\be\nn
\vielb^{x\,a}{}_i{}^k \left( D_x P_{kj} + 2 \sqrt 2 \,\Omega_{kj} K_x + 3 K_x^M \mathcal V_{M\,kj}\right)\;=\; 0\,.
\ee
Hence, multiplying this with an arbitrary coefficient $\alpha$, the gaugino shift matrix~\eqref{gauginoshiftwithonlyK} can be rewritten as
\be\nn
P^a{}_{\!ij}\;=\;
\vielb^{x\,a}{}_i{}^k\left(\alpha\,D_x P_{kj}-\frac{3-8\alpha}{6\sqrt{2}}\,\Omega_{kj}\,K_x-\frac{1-2\alpha}{2}\,K_x^{M}\mathcal{V}_{M\,kj}\right).
\ee
} 

\subsection{Identities in ${\rm su}(2)_\pm$ subspaces}

Our proof of the statement given at the beginning of this appendix crucially relies on the so(5) $\to {\rm so}(4)= {\rm su}(2)_+\oplus {\rm su}(2)_-$ decomposition described in section~\ref{sec:decomposition}. So, as a further preliminary step, we now derive some general identities which require no assumptions and just use the so(5) $\to {\rm so}(4)= {\rm su}(2)_+\oplus {\rm su}(2)_-$ decomposition.

We first introduce the su(2)$_\pm$ components of the curvature of the so(5) connection:
\be\label{defRpm}
(R_\pm^r)_{xy} := \frac{1}{2} R_{xy\,ij} L_\pm^{r\,\, ij} \,=\, 4R_{xy}{}^{\ul{ab}}(L^r_\pm)_{\ul{ab}} \,=\, \vielb_x^{a\ul a}  \vielb_y^{a \ul b} L^r_\pm{}_{\,\ul{ab}} \,.
\ee
Contracting identity \eqref{RRcomplete} with $L_\pm^{r\,ij}L^{s}_{\pm\,kl}$, we find that their product satisfies
\be\label{R+R+}
(R_\pm^{r})_{xy} (R_\pm^{s})^y{}_{z} \,=\, -\frac{1}{16}\delta^{rs}\left( g_{xz} - \vielb_x^{a \underline{a}} \vielb_z^{a \underline{b}} \vct_{\underline{a}}\vct_{\underline{b}}\right) + \frac{1}{4} \epsilon^{rst} (R_\pm^{t})_{xz}\,.
\ee
Note that the first term on the right hand side is a metric constructed with the vielbeine orthogonal to $\vct$ only.

Another relation that will be useful in our proof below is
\be\label{AfRplus=0}
R^r_{\pm}{}_{xy} \vielb^{y\,a\ul{a}}\vct_{\ul{a}} \,=\, 0\,.
\ee
This is easily seen by using \eqref{defRpm} to express $R_{\pm}^r$ in terms of the vielbeine, and noting that in vectorial indices the first equality in~\eqref{eq:XL} reads
$\vct^{\ul{a}}L^r_{\pm}{}_{\ul{ab}}\,=\,0$.

Further useful identities can be deduced starting from the general relation \eqref{DPisRK}.
Let us contract it with $\frac 12 L^r_{\pm}{}^{ij}$ in order to obtain its su$(2)_{\pm}$ components. For the first term on the right hand side we simply use \eqref{defRpm}. For the second term, we notice that
\be
L^r_\pm{}^{ij} R_{xy}{}_{\,i}{}^k\, \mathcal V_{M\, kj} \,=\, \pm\, 3\, L^r_{\pm\,[\ul{ab}}\vct_{\ul{c}]} \vielb_x^{a\ul{a}}\vielb_{y}^{a\ul{b}} \,\mathcal V_M{}^{\ul{c}}\,,
\ee
where we used~\eqref{CurvatureComponents} and then equation
\be\label{epsLX=L}
\epsilon_{\ul{abcde}}L^r_\pm{}^{\ul{cd}}\vct^{\ul e} \,=\, \mp \,2\,L^r_\pm{}_{\ul{ab}},
\ee 
which is just the second equality in \eqref{eq:XL} expressed in vectorial indices.

By further massaging the different terms arising from the antisymmetrization, we eventually arrive at
\begin{equation}\label{DP+=R+KBis}
3\,D_xP_{\pm}^r\,=\,-2\sqrt{2}\,R_{\pm}^r{}_{xy}\big(K^y\,\pm\sqrt{2}\,K^{My}\mathcal{V}_M{}^{\underline{a}}\vct_{\underline{a}}\big) \pm 2\Sigma^{-1} \vielb_x^{a \ul{c}}\vct_{\ul c} L^r_{\pm}{}_{\,\ul{ab}}\, f^{a\ul{ab}}\,,
\end{equation}
where $D_x$ is now an SU(2) covariant derivative: $DP^r_\pm = dP^r_\pm + \frac 12 \epsilon^{rst}\omega^s_\pm P^t_\pm$, with $\omega^s_\pm$ the SU(2)$_\pm$ components of the SO(5) connection.\footnote{A remark is in order about the su(2) components of $DP$. Writing $P_{ij}$ as in~\eqref{expansionPasWQL}, one finds that $\frac 12 L^r_{\pm}{}^{ij}D_xP_{ij} = D_xP^r_\pm $ plus extra terms proportional to $D_x L^r_+$ or $D_x L^r_-$. By working out the covariant derivative $D_x$ of the generators $L$, one sees that in this expression it reduces to a partial derivative, $\partial_x$. Then, working in an SO(5) gauge in which $\vct$ is constant, so that the same holds for the $L$'s (recall footnote~\ref{footnoteSO5gauge}), we obtain $\frac 12 L^r_{\pm}{}^{ij}D_xP_{ij} = D_xP^r_\pm$, which is the result displayed in~\eqref{DP+=R+KBis}. Anyway, we also cross-checked that our results remain the same when no SO(5) gauge-fixing is done and the extra terms are included.}
Substituting $P_{\pm}^r = -iW_{\pm} Q^r_{\pm}$ gives
\begin{equation}\label{intermediatesufficiency}
3i\, \partial_x W_{\pm} Q^r_{\pm} + 3i\, W_{\pm} D_x Q^r_{\pm}\,=\,2\sqrt{2}\, R_{\pm}^r{}_{xy}\big(K^y\,\pm\sqrt{2}\,K^{My}\mathcal{V}_M{}^{\underline{a}}\vct_{\underline{a}}\big) \mp 2\Sigma^{-1} \vielb_x^{a \ul{c}}\vct_{\ul c} L^r_{\pm}{}_{\,\ul{ab}}\, f^{a\ul{ab}}.
\end{equation}
We can decompose this equation in the component parallel to $Q^r_{\pm}$ and its orthogonal part. The former is obtained by contracting with $Q^r_{\pm}$. Noting that from \eqref{Qpmsquare}, \eqref{eq:LL} it follows that 
\be\label{QQis1}
Q_{\pm}^rQ_{\pm}^r= -1
\ee 
and therefore that
\be\label{QDQis0}
Q^r_{\pm} D_x Q^r_{\pm} = 0\,,
\ee 
we obtain
\be\label{dWisQRKcomplete}
-3i\, \partial_x W_{\pm}= 2\sqrt{2}\, Q^r_{\pm} R^r_{\pm xy} \big(K^y\,\pm\sqrt{2}\,K^{My}\mathcal{V}_M{}^{\underline{a}}\vct_{\underline{a}}\big) \mp 2\Sigma^{-1} \vielb_x^{a \ul{c}}\vct_{\ul c} Q^r_{\pm} L^r_{\pm}{}_{\,\ul{ab}} \,f^{a\ul{ab}}\,. 
\ee
The orthogonal components are obtained by projecting \eqref{intermediatesufficiency} with $(\delta^{rs} + Q^r_{\pm}Q^s_{\pm})$, which gives
\begin{equation}\label{otherintermediatestep}
3i\, W_{\pm} D_x Q^r_{\pm}\,=\,\left(\delta^{rs} + Q^r_{\pm}Q^s_{\pm}\right)\left[2\sqrt{2}\,R_{\pm}^s{}_{xy}\big(K^y\,\pm\sqrt{2}\,K^{My}\mathcal{V}_M{}^{\underline{a}}\vct_{\underline{a}}\big) \mp 2\Sigma^{-1} \vielb_x^{a \ul{c}}\vct_{\ul c}L^s_{\pm}{}_{\,\ul{ab}}f^{a\,\ul{ab}}\right].
\end{equation}
Finally, we multiply \eqref{dWisQRKcomplete} by $Q_{\pm}^sR_{\pm}^s$. Using \eqref{R+R+} for the first term on the right hand side, noting that
$\vielb_z^{a \ul{b}}\vct_{\ul b} K^{zM}\mathcal{V}_M{}^{\ul{a}}\vct_{\ul{a}}=0$ identically, and recalling \eqref{AfRplus=0} to see that the last term on the right hand side does not contribute, we arrive at
\begin{equation}\label{RQdWequalto}
\left(g_{xz} - \vielb_x^{a\ul{a}}\vct_{\ul a} \vielb_y^{a \ul{b}} \vct_{\ul b}\right)K^y \,\pm\,\sqrt{2}\,K_x{}^{M}\mathcal{V}_M{}^{\ul{c}}\vct_{\ul{c}} \,=\, -12\sqrt{2}\, i\, Q^r_{\pm}R^r_{\pm}{}_x{}^y\partial_y W_{\pm} \,  .
\end{equation}

Given this set of identities, we are now in the position of proving our statement.

\subsection{Necessity}

We now come to our actual proof.
We first show that conditions \eqref{FloweqFromGaugino}, \eqref{condition_xi_appendix} and \eqref{condition_f_appendix} follow from the gaugino equation \eqref{gauginoEqBIS}, i.e.\ are necessary conditions.
We start contracting~\eqref{gauginoEqBIS} with a vielbein:
\begin{equation}
\vielb^a_{x\,i}{}^l\vielb^a_{y\,l}{}^k\left(i\,\phi^{y\,\prime}\,\delta_k{}^j\gamma_5 + \frac{g}{\sqrt 2}\delta_k{}^j K^y +g\,  K^{My} \mathcal{V}_M{}_k{}^{j}\right)\epsilon_j\,=\,0\,.
\end{equation}
Then we use~\eqref{productvielb} to express the product of the two vielbeine, and constraint \eqref{projectionspinors} to eliminate $\gamma_5$. Moreover, we invoke~\eqref{DPisRK}, \eqref{identitycommutator} to eliminate the terms that involve both the curvature $R$ and the Killing vectors $K$, $K_M$. In this way we arrive at
\begin{equation}
\pm i\,\phi^{y\,\prime}\left(g_{xy}\delta_i{}^k +4R_{xy}{}_{\,i}{}^k\right)Q_+{}_k{}^j\epsilon_j - 3g\,D_xP_i{}^j\epsilon_j +\frac{g}{\sqrt{2}}K_x\,\epsilon_i +2gK_x^M\,\mathcal{V}_M{}_i{}^j\,\epsilon_j\,=\,0\,,
\end{equation}
where the sign choice in the first term is inherited from~\eqref{projectionspinors}.
By projecting with $\Pi_+$ and massaging the different terms (also recalling $\Pi_-\epsilon =0$), we obtain an equation of the form $\left(\alpha\, \delta_i^j + \beta^r L^r_+{}_i{}^j\right)\epsilon_j = 0$. This is equivalent to $\alpha = \beta^r = 0$, as it can be seen by applying $\left(\alpha\, \delta_i^j + \beta^r L^r_+{}_i{}^j\right)$ another time and using property~\eqref{eq:LL} of the su(2) generators. In our case, $\alpha = 0$ corresponds to
\begin{equation}\label{K=R+P+Complete}
K_x+\sqrt{2}\,K_x^M\,\mathcal{V}_M{}^{\underline{a}}\,\vct_{\underline{a}}\,=\, \pm\, 4\sqrt{2}\,i\, g^{-1}\,Q_+^r R_+^r{}_{xy}\phi^{y\,\prime}\,,
\end{equation}
while $\beta^r = 0$ is
\begin{equation}\label{piecewithLcomplete}
3gi\, D_x P^r_+ \,= \, \mp\, g_{xy} \phi^{y\,\prime} \, Q^r_+  \pm 4\,\epsilon^{rst} Q_+^s R_+^t{}_{xy}\phi^{y\,\prime}  \,.
\end{equation}
Since $P_\pm^r=-iW_\pm Q_\pm^r$, the latter is the same as
\be\label{conditionfromgauginoComplete}
3g \, \partial_x W_+ Q^r_+ +  3g\, W_+ D_x Q^r_+ \, = \, \mp \, g_{xy} \phi^{y\,\prime} Q^r_+ \pm\, 4\, \epsilon^{rst} Q_+^s R_+^t{}_{xy} \phi^{y\,\prime}\,.
\ee

Now, eq.~\eqref{K=R+P+Complete} leads to our condition \eqref{condition_xi_appendix}: this is obtained multiplying \eqref{K=R+P+Complete} by $\vct_{\ul a} \vielb_x^{a \ul a}$, recalling \eqref{AfRplus=0}, noting that 
$\,K^{Mx}\,\mathcal{V}_M{}^{\underline{b}}\,\vct_{\underline{b}}\, \vielb_x^{a\ul{a}}\vct_{\ul{a}}\,=\, 0$ identically and then expanding~$K^x$.

On the other hand, recalling \eqref{QQis1}, \eqref{QDQis0}, we see that the component of \eqref{conditionfromgauginoComplete} along $Q^r_+$ is just the wanted flow equation \eqref{FloweqFromGaugino}. Finally, the part of \eqref{conditionfromgauginoComplete} orthogonal to $Q^r_+$ is
\be\label{DxQ}
 3g W_+ D_x Q^r_+ \, = \,  \pm\, 4\,\epsilon^{rst} Q_+^s R_+^t{}_{xy} \phi^{y\,\prime}\,,
\ee
and leads to our constraint~\eqref{condition_f_appendix}. This is seen by contracting the general identity \eqref{otherintermediatestep} with $ \vielb^{x\,a \ul d} \vct_{\ul d}$: the term in \eqref{otherintermediatestep} involving $R^s_+$ vanishes by \eqref{AfRplus=0}, and the same happens for the term involving $D_xQ_+$, once \eqref{DxQ} is invoked.

As an aside, note that~\eqref{DxQ} also implies $\phi^{x\,\prime} D_x Q^r_+ =0$, which is one of the constraints we found while analysing the gravitino equation in section~\ref{AnalysisGravityMult}. Therefore this constraint is automatically satisfied once the gaugino equation is solved.

\newpage

\subsection{Sufficiency}

It remains to prove that the flow equation \eqref{FloweqFromGaugino} together with constraints \eqref{condition_xi_appendix}, \eqref{condition_f_appendix} are sufficient conditions for the gaugino equation~\eqref{gauginoEqBIS}. To see this, we first show that they imply \eqref{K=R+P+Complete}, as well as the following relation involving the component of the so(5)-valued matrix $f^{a\ul{ab}}$ parallel to $Q^r_+\,$:
\be\label{constraintcomplete}
\mp i\, \phi^{x\,\prime}  \vielb_x^{a\ul{a}}\vct_{\ul{a}} \,=\, 2g\,\Sigma^{-1} Q^r_+ L_+^r{}_{\ul{ab}} f^a{}^{\ul{ab}}\,.
\ee

Eq.~\eqref{K=R+P+Complete} is immediately obtained from identity \eqref{RQdWequalto} by using \eqref{FloweqFromGaugino}, \eqref{condition_xi_appendix}.
On the other hand, contracting identity \eqref{dWisQRKcomplete} with $\vct_{\ul{a}} \vielb^{x\, a \ul{a}}$ and recalling~\eqref{AfRplus=0} yields
\be
3i\, \partial_x W_+ \vielb^{x\, a \ul{a}} \vct_{\ul{a}} \,=\, 2\,\Sigma^{-1} Q^r_+ L^r_+{}_{\ul{ab}}f^{a\ul{ab}}\,,
\ee
which by using \eqref{FloweqFromGaugino} becomes the wanted relation \eqref{constraintcomplete}.

We next show that, together with constraints \eqref{projectionspinors} and \eqref{condition_f_appendix}, the relations just obtained satisfy the gaugino equation~\eqref{gauginoEqBIS}.
Using constraint \eqref{projectionspinors} to eliminate $\gamma_5$, as well as \eqref{K=R+P+Complete} to eliminate $K^x$, eq.~\eqref{gauginoEqBIS} becomes
\begin{equation}\label{gauginoalmostdonecomplete}
\mp i\,\phi^{x\,\prime} \left(\vielb^a_{x\,i}{}^k Q_{+k}{}^j + 4\, \vielb^{y\,a}{}_i{}^j R^r_+{}_{yx}Q^r_+ \right) \epsilon_j  +  g\, K^{Mx} \vielb^a_{x\,i}{}^k \left( \delta_k^j\,\mathcal{V}_M{}^{\underline{a}}\vct_{\underline{a}} -  \mathcal{V}_M{}_k{}^{j} \right)\epsilon_j \,=\,0\,.
\end{equation}
Let us elaborate the first and the second term separately.
Expressing $R_+^r$ as in \eqref{defRpm}, one can check that 
\be\vielb^a_{x\,i}{}^k Q_{+k}{}^j + 4\, \vielb^{y\,a}{}_i{}^j R^r_+{}_{yx}Q^r_+\,=\, 
Q_+{}_i{}^k \vielb^a_{x\,k}{}^j \,.
\ee
Also noting that
\be
Q_+ \vielb_x^a \,=\, Q_+ \Pi_+ \vielb_x^a \,=\, \vielb_x^{a\ul{a}} \vct_{\ul a} Q_+ + Q_+ \vielb_x^a\, \Pi_-\,,
\ee  
and that the last term annihilates $\epsilon$, we obtain for the first term of \eqref{gauginoalmostdonecomplete}:
\bea
\mp i\,\phi^{x\,\prime} \left(\vielb^a_{x\,i}{}^k Q_{+k}{}^j + 4\, \vielb^{y\,a}{}_i{}^j R^r_+{}_{yx}Q^r_+ \right) \epsilon_j 
&=& \mp i \, \phi^{x\,\prime}  \vielb_x^{a\ul{a}} \vct_{\ul a} Q_{+i}{}^j\epsilon_j \nonumber \\ [2mm]
&=&  2g\,\Sigma^{-1} Q^r_+ L_+^r{}_{\ul{ab}} f^a{}^{\ul{ab}}\,Q_{+i}{}^j\epsilon_j\,,
\eea
where in the last equality we used condition \eqref{constraintcomplete}.
On the other hand, the second term of \eqref{gauginoalmostdonecomplete} can be rewritten as
\bea
g\, K^{Mx} \vielb^a_{x\,i}{}^k \left( \delta_k^j\,\mathcal{V}_M{}^{\underline{a}}\,\vct_{\underline{a}} - \mathcal{V}_M{}_k{}^{j} \right)\epsilon_j 
&=&
g\, K^{Mx} \vielb^a_{x\,i}{}^k \left[ \vct_k{}^l \mathcal{V}_M{}_l{}^{j} -  2\mathcal{V}_M{}_k{}^{l} \Pi_-{}_l{}^j \right]\epsilon_j \nonumber\\ [2mm]
&=& g\, K^{Mx} \vielb^a_{x\,i}{}^k  \vct_k{}^l \mathcal{V}_M{}_l{}^{j}\epsilon_j\nonumber\\ [2mm]
&=& -\frac 14 g \,\Sigma^{-1} f^{a\ul{ab}} \epsilon_{\ul{abcde}}\vct^{\ul{e}}\, \Gamma^{\ul{cd}}{}_i{}^j \epsilon_j\,,
\eea
where the last equality is obtained by expanding $K^{Mx}$.
So \eqref{gauginoalmostdonecomplete} becomes
\be\label{fVVVetc}
f^{a\ul{ab}} \left( 8\,Q_{+}^r L^r_+{}_{\ul{ab}} Q_+{}_i{}^j - \epsilon_{\ul{abcde}}\vct^{\ul e} \Gamma^{\ul{cd}}{}_i{}^j\right) \epsilon_j \,=\,0\,.
\ee
Let us elaborate the second term in parenthesis. We can write
\bea
 \epsilon_{\ul{abcde}}\vct^{\ul e}\, \Gamma^{\ul{cd}}{}_i{}^j \epsilon_j 
&=& \epsilon_{\ul{abcde}}\vct^{\ul e} \left(\Gamma^{\ul{cd}}\Pi_+\Pi_+\right)_i{}^j\epsilon_j \nonumber \\[2mm]
&=&  \epsilon_{\ul{abcde}}\vct^{\ul e}\left(\Pi_+\Gamma^{\ul{cd}}\Pi_+\right)_i{}^j \epsilon_j \nonumber \\[2mm]
&=& 4\,\epsilon_{\ul{abcde}}\vct^{\ul e} L^r_+{}^{\ul{cd}}\, L^r_+{}_i{}^j \epsilon_j  \nonumber \\[2mm]
&=& -8\,L^r_+{}_{\ul{ab}} L^r_+{}_i{}^j\epsilon_j\,,
\eea
where in the first equality we used $\epsilon = \Pi_+\epsilon$ twice, in the second we noted that $\vct \equiv \vct_{\ul{f}}\Gamma^{\ul{f}}$ commutes with $\epsilon_{\ul{abcde}}\vct^{\ul e} \Gamma^{\ul{cd}}$, 
in the third we used the completeness relation \eqref{completenessL}, and in the last we recalled~\eqref{epsLX=L}.
So \eqref{fVVVetc} becomes
\be
f^{a\ul{ab}}L^r_{+\ul{ab}} \left(\delta^{rs} + Q_{+}^r Q^s_+  \right)L^s_+{}_i{}^j \epsilon_j \,=\,0\,,
\ee
which is satisfied by \eqref{condition_f_appendix}.

This concludes our proof that the gaugino equation~\eqref{gauginoEqBIS} is equivalent to \eqref{FloweqFromGaugino}, \eqref{condition_xi_appendix} and \eqref{condition_f_appendix} once the constraint \eqref{projectionspinors} on the spinor is taken into account and the choice of $W_+$ is made.


\end{document}